# Analytical modeling of micelle growth. 3. Electrostatic free energy of ionic wormlike micelles – effects of activity coefficients and spatially confined electric double layers

Krassimir D. Danov [a], Peter A. Kralchevsky [a,*], Simeon D. Stoyanov [b,c,d],
Joanne L. Cook [e], Ian P. Stott [e]

[a] *Department of Chemical and Pharmaceutical Engineering, Faculty of Chemistry and Pharmacy, Sofia University, Sofia 1164, Bulgaria*

[b] *Unilever Research & Development Vlaardingen, 3133AT Vlaardingen, The Netherlands*

[c] *Laboratory of Physical Chemistry and Colloid Science, Wageningen University, 6703 HB Wageningen, The Netherlands*

[d] *Department of Mechanical Engineering, University College London, WC1E 7JE, UK*

[e] *Unilever Research & Development Port Sunlight, Bebington CH63 3JW, UK*

ORCID Identifiers: Krassimir D. Danov: 0000-0002-9563-0974 ; Peter A. Kralchevsky: 0000-0003-3942-1411 ; Simeon D. Stoyanov: 0000-0002-0610-3110

ABSTRACT

*Hypotheses*: To correctly predict the aggregation number and size of wormlike micelles from ionic surfactants, the molecular-thermodynamic theory has to calculate the free energy per molecule in the micelle with accuracy better than 0.01 *kT*, which is a serious challenge. The problem could be solved if the effects of mutual confinement of micelle counterion atmospheres, as well as the effects of counterion binding, surface curvature and ionic interactions in the electric double layer (EDL), are accurately described.

*Theory*: The electric field is calculated using an appropriate cell model, which takes into account the aforementioned effects. Expressions for the activity coefficients have been used, which vary across the EDL and describe the electrostatic, hard sphere, and specific interactions between the ions. New approach for fast numerical calculation of the electrostatic free energy is developed.

*Findings*: The numerical results demonstrate the variation of quantities characterizing the EDL of cylindrical and spherical micelles with the rise of electrolyte concentration. The effect of activity coefficients leads to higher values of the free energy per surfactant molecule in the micelle as compared with the case of neglected ionic interactions. The results are essential for the correct prediction of the size of wormlike micelles from ionic surfactants. This study can be extended to mixed micelles of ionic and nonionic surfactants for interpretation of the observed synergistic effects.

* Corresponding author. Tel.: +359 2 962 5310
  E-mail address: pk@lcpe.uni-sofia.bg (P.A. Kralchevsky)

# 1. Introduction

The present series of papers is devoted to the development of a molecular thermodynamic theory of the growth of wormlike micelles (WLM) that is able to predict their mean aggregation number in agreement with the experiment. For this goal, in Ref. [1] we presented a detailed review on the state of the art with some new results concerning the micelle chain-conformation free energy and the procedure for comparison theory and experiment. Insofar as a comprehensive review has been already published [1], here we will focus mostly on papers that are closely related to the subject of the present article, viz. molecular-thermodynamic theory of WLM from *ionic* surfactants.

In Ref. [1], it was demonstrated that to predict the WLM mean mass aggregation number, $n_M$, the theory should be able to calculate the excess free energy per surfactant molecule in the micelle endcaps (the so-called scission energy) with high precision, which has to be better that 0.01 $k_B T$ – see Section 5 of the present article. This is the main challenge, which stimulated us to construct a theoretical description of enhanced precision in the subsequent papers of this series, viz. Refs, [2,3,4] and the present article.

In Ref. [2], the analytical mean-field theory of chain conformation free energy of the micellar hydrophobic core was extended to the case of mixed micelles. It was established that the mixing of surfactants with different hydrocarbon chainlengths is always synergistic.

In Ref. [3], a thermodynamic expression for the scission energy of mixed micelles was derived. The molecular-thermodynamic theory was compared with available experimental data for the aggregation number $n_M$ for nonionic surfactant micelles and agreement theory-experiment was achieved without using any adjustable parameters.

Here, our goal is to extend the theory to ionic surfactant solutions, which implies calculation of the micelle electrostatic free energy with enhanced precision, removing approximations used in previous studies.

Ninham et al. [5,6] developed an elegant theory based on integration of the relation between surface charge and surface potential. Analytical formula for micelle electrostatic free energy was derived at the cost of several approximations [5,6]: (i) Infinite electric double layer (EDL) around each micelle; (ii) ideal electrolyte solution (i.e., ionic activity coefficients $\gamma_i \equiv 1$); (iii) Use of approximated evaluation of the free-energy integral and truncated series expansions to take into account the curvature effect, and (iv) neglected effect of counterion binding. Nagarajan & Ruckenstein [7] incorporated this model of micelle electric energy in their theory of micellization.

In subsequent studies, the theory from [5,6] was upgraded to avoid a part of the used approximations or simplifying assumptions. Alargova et al. [8,9] and Srinivasan and Blankschtein [10] demonstrated that the effect of counterion binding has to be taken into account in order to achieve agreement between theory and experiment, especially in the case of multivalent counterions. Koroleva and Victorov [11] took into account the effect of the finite size of the ions by using the Boublik−Mansoori−Carnahan−Starling−Leland (BMCSL) equation for a mixture of hard spheres of different radii [12-14]. Note, however, that all these studies are still using some of the simplifying assumptions adopted in Refs. [5,6].

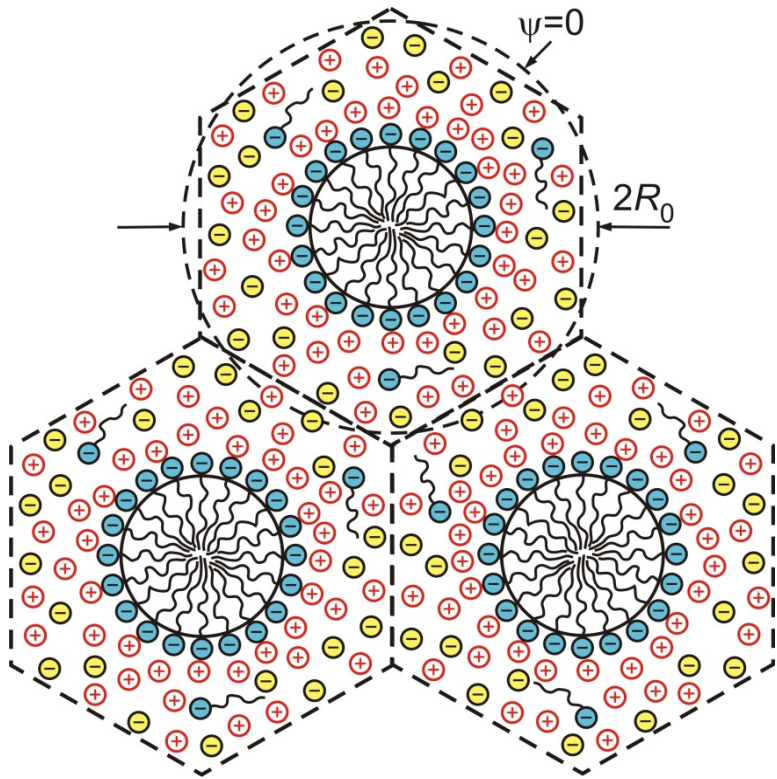

**Fig. 1.** Sketch of the used cell model of a micellar solution. The electric double layer around each micelle (cylindrical or spherical) is closed in a cell of outer radius $R_0$. At the outer boundary of each cell, both the electric potential and field are assumed to be zero: $\psi = 0$ and $E = 0$.

Here we develop a different approach to the calculation of micelle electrostatic free energy, which avoids using all aforementioned approximations and simplifying assumptions and meets the requirement for enhanced accuracy needed for the theoretical prediction of WLM growth. In particular, the approximate assumption that the micelle electrostatic potential decays at infinity is removed. Instead, the electric field is calculated using an appropriate cell model, which takes into account the mutual spatial confinement of the EDLs of the neighboring micelles (Fig. 1). The model is based on the Poisson equation and the mass balances of all ionic species in the solution. Similar (but not identical) cell models have been previously used to quantify the electrostatic interactions in micellar solutions [15] and colloidal dispersions [16-18].

Usually, wormlike micelles from ionic surfactants are formed at high salt concentrations, in the range 0.4 – 4 M [19-25]. For this reason, the effect of ionic activity coefficients, $\gamma_i$, becomes important. In our model, $\gamma_i$ varies across the micellar EDL as a function of the local ionic concentrations. For such detailed model, the semiempirical approach by Pitzer and other authors [26-31] to the quantitative description of activity coefficient is inappropriate, because it has been designed for uniform electrolyte solutions. Here, theoretical expressions for $\gamma_i$ are used, which take into account (i) electrostatic [32,33]; (ii) hard-sphere [12-14] and (iii) specific interactions [34] between the ions, and exactly describe the experimental dependencies of $\gamma_\pm = (\gamma_+\gamma_-)^{1/2}$ on the salt concentration for uniform solutions [35]. Furthermore, to describe the electrostatic potential and the ionic distributions in the EDL, we combine the Poisson equation with the equation for electrochemical equilibrium (with $\gamma_i \neq 1$), rather than with the conventional Boltzmann equation, which presumes $\gamma_i = 1$. In this respect, the present study is different from the Poisson-Boltzmann (PB) model used in many preceding studies.

The upgrade of theory with additional effects (and especially, with $\gamma_i = \gamma_i(r) \neq 0$) demands development of a new approach to the calculation of micelle free energy. The old one [5,6] would lead to many times repeated numerical solutions of the electrostatic boundary-value problem, which makes it practically unusable for our goal. To overcome this problem, we derived a different (but equivalent) expression for the micelle free energy, which allows using one-time numerical solution of the boundary-value problem. At that, the effect of micelle surface curvature on the EDL is taken into account exactly (without any truncated series expansions), and the effect of counterion binding has been described via the Stern isotherm [36].

The paper is organized as follows. Section 2 describes the derivation of the new appropriate expression for micelle electrostatic free energy, which is obtained without using the Boltzmann equation. Section 3 presents the cell model and the way for solving the arising electrostatic boundary-value problem. The computational procedure is described in SI Appendix F (SI = Supplementary Information). Section 4 is devoted to the theoretical model of ionic activity coefficients and to determining the parameters of this model from experimental data for uniform electrolyte solutions. Section 5 presents numerical results and discussion.

The next step is to compare the developed theory with available sets of experimental data for the mean mass aggregation number of WLM, $n_M$, vs. the salt concentration and temperature, $T$, for both anionic and cationic surfactants [19-25]. This is the subject of the next part of this series, Ref. [4]. There, the results of the present paper are utilized to calculate the electrostatic component of WLM scission energy, which is combined with the other three free-energy components related to the interfacial tension, headgroup steric repulsion and surfactant chain conformations within the micelle. Finally, the resulting model is tested against the experiment.

Some aspects of the present study, such as the cell model; the new approach for solving the electrostatic boundary-value problem, and the developed theoretical description of activity coefficients could find applications for other colloidal systems with developed EDLs (not necessary surfactant micelles), such as the particle interactions in dispersions and porous media in both quasi-equilibrium [37,38] and electrokinetic [38-40] phenomena.

**2. Electrostatic free energy of colloidal dispersions**

Here, our goal is to derive an expression for the electrostatic free energy of colloidal dispersions, including surfactant micelles, which is convenient for applications in the case of high ionic strengths, at which the effect of activity coefficients has to be taken into account. Moreover, this expression exactly describes the effect of particle (e.g. micelle) surface curvature, without using any truncated series expansions.

*2.1. General equations for the EDL around charged colloids*

As already mentioned, we will consider each electrically charged micelle (spherical, cylindrical, discoidal, etc.), to be contained in a cell of outer boundary at which the electric field and potential are supposed to be equal to zero (Fig. 1). The same approach is applicable also to the EDL of charged emulsion drops, gas bubbles or solid beads in colloidal dispersions. In the special case of diluted dispersions, one could set the outer boundary of the cell at infinity.

The *electrostatic energy*, $U_{el}$, can be described as the field energy in the solution, or, alternatively as the energy of the bulk and surface charges in the local potential field [41,42]:

$$U_{el} = \frac{\varepsilon \varepsilon_0}{2} \int_V E^2 \, dV = \frac{1}{2} \int_V \rho_b \psi \, dV + \frac{1}{2} \int_A \rho_s \psi_s \, dA \tag{2.1}$$

Here, $\varepsilon$ is the dielectric constant of solution; $\varepsilon_0$ is the permittivity of vacuum; $\rho_b$ and $\rho_s$ are the bulk and surface electric charge densities; $\psi$ and $\psi_s$ are the bulk and surface electric potentials; $\mathbf{E} = -\nabla \psi$ is the vector of electric field; $E^2 = \mathbf{E} \cdot \mathbf{E}$; $dV$ and $dA$ are volume and surface elements. The equivalence of the two presentations of $U_{el}$ in Eq. (2.1) can be proven by means of the Gauss's divergence theorem – see SI Appendix A. For our goal, it is convenient to represent the expression for $U_{el}$ in another equivalent form:

$$U_{el} = \int_V (-\frac{\varepsilon_0 \varepsilon}{2} E^2 + \rho_b \psi) \, dV + \int_A \rho_s \psi_s \, dA \tag{2.2}$$

In view of Eq. (2.2), the *free energy* of the EDL can be expressed in the form:

$$F_{EDL} = \int_V (-\frac{\varepsilon_0 \varepsilon E^2}{2} + \rho_b \psi + f_b) \, dV + \int_A \rho_s \psi_s \, dA \tag{2.3}$$

where $f_b$ is the bulk density of the non-electrostatic contribution to the free energy.

To find an expression for $f_b$, we will use the classical Gibbs approach and will consider a nonuniform system as composed of a large number of small domains, such that in each of them the system can be treated as uniform; see e.g. Ref. [43]. The Gibbs fundamental equation for such domain reads:

$$dF = -S \, dT - p \, dV + \sum_{i=1}^{m} \mu_i \, dN_i \tag{2.4}$$

Here, $F$ is free energy; $T$ is temperature; $p$ is pressure, $V$ is volume; $\mu_i$ is chemical potential, $N_i$ is number of molecules, and the summation is over all components, $1 \leq i \leq m$. Because, the considered domain is supposed to be uniform, the integration of Eq. (2.4) over the volume $V$ yields:

$$F = -pV + \sum_{i=1}^{m} \mu_i N_i \tag{2.5}$$

By definition, $f_b = F/V$ and then Eq. (2.5) acquires the form

$$f_b = -p + \sum_{i=1}^{m} \mu_i c_i \tag{2.6}$$

where $c_i = N_i/V$ are the local concentrations ($1 \leq i \leq m$). Furthermore, by substituting Eq. (2.5) in Eq. (2.4), one derives the known Gibbs-Duhem equation:

$$\mathrm{d}p = s_b\,\mathrm{d}T + \sum_{i=1}^{m} c_i\,\mathrm{d}\mu_i \tag{2.7}$$

where $s_b = S/V$ is the local density of entropy. Substituting Eqs. (2.6) in Eq. (2.3), we obtain:

$$F_{EDL} = \int_V \left(-\frac{\varepsilon_0 \varepsilon E^2}{2} - p + \sum_{i=1}^{m} c_i \mu_i^{el}\right) \mathrm{d}V + \int_A \rho_s \psi_s\,\mathrm{d}A \tag{2.8}$$

where we have used the definitions of the bulk charge density and electrochemical potential:

$$\rho_b = \sum_{i=1}^{m} q_i c_i \tag{2.9}$$

$$\mu_i^{el} = \mu_i + q_i \psi = \mathrm{const.} \tag{2.10}$$

and $q_i$ is the charge of the respective molecule. The constancy of $\mu_i^{el}$ follows from the condition for electrochemical equilibrium across the EDL [42,43]. Under isothermal conditions ($T$ = const.), with the help of Eqs. (2.7), (2.9) and (2.10) we obtain:

$$\mathrm{d}p = \sum_{i=1}^{m} c_i\,\mathrm{d}\mu_i = \sum_{i=1}^{m} c_i\,\mathrm{d}(\mu_i + q_i\psi) - \sum_{i=1}^{m} q_i c_i\,\mathrm{d}\psi = -\rho_b\,\mathrm{d}\psi \tag{2.11}$$

where the constancy of the electrochemical potential has been used. The integration of Eq. (2.11) yields:

$$p - p_0 = -\int_0^{\psi} \rho_b(\tilde{\psi})\,\mathrm{d}\tilde{\psi} \tag{2.12}$$

where $p_0$ is the pressure in the region with $\psi = 0$ and $\tilde{\psi}$ is an integration variable. Using Eq. (2.12), we can present $F_{EDL}$ in Eq. (2.8) as a sum of mechanical, chemical and electrostatic contributions, $F_{EDL} = F_{mech} + F_{chem} + F_{el}$, where

$$F_{mech} = -p_0 V; \quad F_{chem} = \sum_{i=1}^{m} \mu_{i,0} N_i^{EDL} \tag{2.13}$$

$$F_{el} = \int_V \left[-\frac{\varepsilon_0 \varepsilon E^2}{2} + \int_0^{\psi} \rho_b(\tilde{\psi})\,\mathrm{d}\tilde{\psi}\right]\mathrm{d}V + \int_A \rho_s \psi_s\,\mathrm{d}A \tag{2.14}$$

Here, $V$ is the volume of the EDL; $\mu_{i,0}$ is the chemical potential in the region with $\psi = 0$; in view of Eq. (2.10), $\mu_{i,0} = \mu_i^{el}$, and $N_i^{EDL}$ is the number of molecules of the respective component in the EDL:

$$N_i^{EDL} = \int_V c_i\,\mathrm{d}V \tag{2.15}$$

Using Eq. (2.1) and the Poisson equation, $\varepsilon\varepsilon_0 \nabla \cdot \mathbf{E} = \rho_b$, one can eliminate the term with $\rho_s \psi_s$ and bring Eq. (2.14) in another equivalent form (see SI Appendix B):

$$F_{el} = \varepsilon\varepsilon_0 \int_V [\frac{E^2}{2} - \psi \nabla \cdot \mathbf{E} + \int_0^\psi (\nabla \cdot \mathbf{E}) d\tilde{\psi}] dV \tag{2.16}$$

*2.2. Discussion*

It is very important to note that Eq. (2.16) was derived without using any specific expression for the chemical potentials $\mu_i$. This means that Eq. (2.16) can be used with any expression for the activity coefficient, $\gamma_i$. Overbeek [42] derived Eq. (2.16) by using the Boltzmann equation, which means that he was working in the special case with $\gamma_i = 1$. The electrochemical potential can be expressed in the following general form:

$$\mu_i^{el} = \mu_i^o + k_B T \ln(\gamma_i c_i) + q_i \psi \tag{2.17}$$

where $\mu_i^o$ is standard chemical potential. Then, using the uniformity of the electrochemical potential and setting $\mu_i^{el} = \mu_{i,0}$ we obtain $k_B T \ln(\gamma_i c_i) + q_i \psi = k_B T \ln(\gamma_{i,0} c_{i,0})$, which is equivalent to

$$\gamma_i c_i = \gamma_{i,0} c_{i,0} \exp\left(-\frac{q_i \psi}{k_B T}\right) \tag{2.18}$$

Here, $c_{i,0}$ and $\gamma_{i,0}$ are the concentrations and activity coefficients in the region with $\psi = 0$. The conventional Boltzmann equation corresponds to $\gamma_i = \gamma_{i,0} = 1$.

Another frequently used expression for the electrostatic free energy, derived by Verwey and Overbeek [44], was applied to micellar systems in the framework of the assumption $\gamma_i = \gamma_{i,0} = 1$ [5,6]:

$$F_{el} = \int_A (\int_0^{\rho_s} \psi_s d\tilde{\rho}_s) dA \tag{2.19}$$

where $\tilde{\rho}_s$ is the surface charge density as an integration variable. Eq. (2.19) allows one to calculate $F_{el}$ if a relation between the surface potential and charge, $\psi_s = \psi_s(\rho_s)$, is available. In view of the way of its derivation, Eq. (2.19) is applicable to symmetrical systems (sphere, cylinder, plane), for which the electric field and potential depend only on the magnitude of position vector, $r = |\mathbf{r}|$: $E = E(r)$, $\psi = \psi(r)$, and consequently, the function $E = E(\psi)$ is also defined. Overbeek [42] derived Eq. (2.19) from Eq. (2.16) assuming $E = E(\psi)$, but without using the Boltzmann equation (see also SI Appendix C). Insofar as Eq. (2.16) holds in the general case with $\gamma_i \neq 1$, it follows that Eq. (2.19) is also applicable with $\gamma_i \neq 1$ for symmetric systems, where $E = E(\psi)$.

In the case with $\gamma_i \neq 1$ and counterion binding at the charged surface, the electrostatic boundary-value problem has to be solved numerically by using also iterations (see below). In such a case, the integration in Eq. (2.19) has to be also carried out numerically, calculating $\psi_s = \psi_s(\rho_s)$ many times, at each step of the numerical integration. In our case, the application of such computational procedure would be so heavy and slow that it becomes difficult to use, and the accumulation of computational errors would be difficult to assess. For this reason, in Section 2.3 we bring the general expression for $F_{el}$ in another equivalent form, which allows one to calculate $F_{el}$ with one-time numerical solution of the electrostatic boundary-value problem.

*2.3. $F_{el}$ in terms of the electrostatic surface pressure*

Here, we will consider the special case of symmetrical system (spherical, cylindrical or planar uniformly charged surface), in which the electric field is directed normal to the charged surface and depends on the distance to it, $E = E(r)$. In such a case, the contribution of the EDL to the surface pressure can be presented in the form of a surface excess [45,46]:

$$\pi_{el} = \int_0^{R_0} (P_T - p_0) dr \quad \text{(plane)} \tag{2.20}$$

$$\pi_{el} = \frac{1}{R_{el}^s} \int_{R_{el}}^{R_0} (P_T - p_0) r^s dr \quad \text{(cylinder, sphere)} \tag{2.21}$$

where $P_T$ is the tangential (with respect to the surface) component of the Maxwell electric pressure tensor; the integration is carried out across the EDL; the surface charges are located at $r = 0$ for the planar surface and at $r = R_{el}$ for the cylindrical and spherical surface; $r = R_0$ is the outer boundary of the cell, where $\psi = 0$ (see Fig. 1). Here and hereafter, $s = 1$ for cylindrical geometry and $s = 2$ for spherical geometry.

The general expression for the Maxwell electric pressure tensor reads [47]:

$$P_{ik} = (p + \frac{\varepsilon \varepsilon_0}{2} E^2) \delta_{ik} - \varepsilon \varepsilon_0 E_i E_k \quad (i,k = 1,2,3) \tag{2.22}$$

where $\delta_{ik}$ is the Kronecker delta symbol (the unit matrix) and $E_i$ is the $i$-th component of the electric field **E**; $p$ is the local hydrostatic pressure. In the considered case of symmetric system, **E** is directed normal to the charged surface, so that $P_T = p + \varepsilon\varepsilon_0 E^2/2$. Then,

$$\pi_{el} = \int_0^{R_0} (\frac{\varepsilon\varepsilon_0}{2} E^2 + p - p_0) dr \quad \text{(plane)} \tag{2.23}$$

$$\pi_{el} = \frac{1}{R_{el}^s} \int_{R_{el}}^{R_0} (\frac{\varepsilon\varepsilon_0}{2} E^2 + p - p_0) r^s dr \quad \text{(cylinder, sphere)} \tag{2.24}$$

In view of Eqs. (2.12), (2.23) and (2.24), for the considered case of symmetric system Eq. (2.14) can be presented in the form

$$F_{el} = A(\rho_s \psi_s - \pi_{el}) \tag{2.25}$$

where, as usual, $A$ is the surface area and $\pi_{el}$ is the electrostatic surface pressure given by Eq. (2.23) or (2.24).

From computational viewpoint, it is convenient to eliminate the term $(p - p_0)$ in the expression for $\pi_{el}$. For this goal, we will use the Poisson equation. For cylindrical ($s = 1$) and spherical ($s = 2$) geometry, this equation reads:

$$\varepsilon\varepsilon_0 (\frac{d^2 \psi}{dr^2} + \frac{s}{r} \frac{d\psi}{dr}) = -\rho_b \tag{2.26}$$

Let us multiply Eq. (2.26) by $d\psi/dr$; integrate from $r$ to $R_0$ and use Eq. (2.12):

$$-\varepsilon\varepsilon_0 \frac{E^2}{2} + \varepsilon\varepsilon_0 \int_r^{R_0} \frac{s}{\tilde{r}} E^2 d\tilde{r} = p_0 - p \tag{2.27}$$

Substituting $(p_0 - p)$ from Eqs. (2.27) into (2.24) and integrating by parts, we obtain (see SI Appendix D):

$$\pi_{el} = \frac{\varepsilon\varepsilon_0}{s+1} \int_{R_{el}}^{R_0} [(\frac{r}{R_{el}})^s + \frac{sR_{el}}{r}] E^2 dr \quad \text{(cylinder, sphere)} \tag{2.28}$$

For planar geometry, for which there is no integral term in Eq. (2.27), the final formula for $\pi_{el}$ reads:

$$\pi_{el} = \varepsilon\varepsilon_0 \int_0^{R_0} E^2 dr \quad \text{(plane)} \tag{2.29}$$

Note that Eq. (2.28) can be derived directly from a general expression for the surface tension of a curved interface obtained in Ref. [46] – see SI Appendix E.

In summary, the electrostatic free energy of the diffuse EDL per unit surface area, $F_{el}$, is given by Eq. (2.25), where the electrostatic surface pressure $\pi_{el}$ is given by Eq. (2.28) or (2.29). The solution of the electrostatic boundary-value problem (see Section 3) yields $\rho_s$, $\psi_s$, and $E(r)$, and then from Eqs. (2.28) or (2.29) one determines $\pi_{el}$; see SI Appendix F. This procedure for calculation of $F_{el}$ can be used with any expression for the activity coefficient, $\gamma_i$. The curvature effects are taken into account exactly, without using any truncated series expansions as in Ref. [5].

*2.4. Surface electrostatic free energy density*

Let us consider monovalent surface ionized groups and monovalent counterions. (In the case of surfactant micelles, this means that both surfactant and salt are 1:1 electrolytes and the counterions due to the surfactant and salt are the same.) The valence of the surface ionized groups will be denoted $z_1$, so that the valence of the counterions is $z_2 = -z_1$; $z_1 = \pm 1$.

The field of the surface ion creates a potential well, i.e. adsorption site, where the counterions might bind. (In addition, there could be also binding energy of non-electrostatic origin.) The bound counterions form the Stern layer. Let $\Gamma_1$ and $\Gamma_2$ be the surface densities of ionized surface groups and bound counterions, respectively. Then, $\rho_s = z_1 e \Gamma_1 + z_2 e \Gamma_2$ (with $e$ being the elementary charge) and using Eq. (2.25) we obtain:

$$\tilde{f}_{el} = \frac{F_{el}}{N_1} = \frac{1}{\Gamma_1}[(z_1 e \Gamma_1 + z_2 e \Gamma_2)\psi_s - \pi_{el}] = (1-\theta)z_1 e \psi_s - \pi_{el}/\Gamma_1 \qquad (2.30)$$

where $\tilde{f}_{el}$ is electric free energy per unit surface charge (in the case of ionic surfactant micelle – per surfactant molecule in the micelle); $N_1$ is the number of surface ionized groups and $N_1/A = \Gamma_1$ is their density; $\theta = \Gamma_2/\Gamma_1$ is the occupancy of the Stern layer.

Note that $\tilde{f}_{el}$ takes into account only the contribution of the *diffuse part* of the EDL. The total electrostatic free energy per surface charge (per surfactant molecule in the micelle) contains contributions from both the diffuse EDL and the Stern layer:

$$f_{el} = \tilde{f}_{el} + \theta z_1 e \psi_s + k_B T \ln(1-\theta) = z_1 e \psi_s - \pi_{el}/\Gamma_1 + k_B T \ln(1-\theta) \qquad (2.31)$$

where the last term accounts for the configurational free energy of the counterions in the Stern layer (Indeed, for $\theta = 0$, i.e., no Stern layer, the two terms added to $\tilde{f}_{el}$ vanish.) The derivation of Eq. (2.31), which is based on extensive thermodynamic considerations, can be found in the next part of this series, Ref. [4].

## 3. Cell model and solution of the electrostatic boundary-value problem

The model developed in Refs. [5,6] and applied by other authors [7-11] assume that the electric field of each separate micelle decays at infinity, where the existence of uniform and electroneutral solution is assumed. This model is appropriate for diluted micellar solutions, near the CMC, where the distance between the micelles is significantly greater than the Debye length. However, in more concentrated surfactant solutions, the electric double layers around the micelles overlap and the solution around a given micelle becomes nonuniform and locally non-electroneutral. (The non-uniformity is related to the fact that the micelles are macroions – particles with hydrocarbon core.) In such a case, the adequate physical model is the cell model [15]. In this model, the electrostatic boundary-value problem is solved for a cell that contains the micelle (or another charged colloidal particle) and its counterion atmosphere; see Fig. 1. In the case of spherocylindrical micelle, cylindrical and spherical cells have been used, respectively, for the cylindrical part of the micelle and its endcaps. The outer cell radius, $R_0$, which is different for the cylinder and the endcaps, is determined in the course of the solution of the electrostatic boundary-value problem, as explained below. The procedure is applicable also to charged spherical micelles.

We will consider ionic surfactant and salt, which are 1:1 electrolytes. It is assumed that the counterions due to surfactant and electrolyte are the same (e.g. Na$^+$ ions for SDS and NaCl). In such a case, the bulk charge density, $\rho_b$, and the dimensionless surface potential, $\Psi$, can be presented in the form:

$$\rho_b = q(c_1 - c_2 + c_3) \quad \text{and} \quad \Psi \equiv \frac{q\psi}{k_B T} > 0 \tag{3.1}$$

As before, $q$ is the electric charge of the surfactant ion ($q = +e$ for cationic surfactant and $q = -e$ for anionic one); $c_1$, $c_2$ and $c_3$ are, respectively, the local bulk concentrations of surfactant ions, counterions and coions due to the added salt.

The input parameters are the total concentrations of surfactant and salt, $C_1$ and $C_3$, which have been dissolved by the experimentalist to prepare the solution. The total concentration of counterions is $C_2 = C_1 + C_3$. Other input parameters are the radius $R_{el}$ of the surface, where the surface charges are located; the number of surfactant ionized headgroups per unit area of micelle surface, $\Gamma_1$, and the concentration of surfactant ions $c_{1,0}$ in the region with $\psi = 0$. (The equilibrium values of $R_{el}$, $\Gamma_1$ and $c_{1,0}$ are determined when the total free energy is minimized to find the equilibrium state of the micelle [4].)

In view of Eq. (3.1), we can represent Eq. (2.26) in the form:

$$\frac{1}{r^s}\frac{\partial}{\partial r}(r^s \frac{\partial \Psi}{\partial r}) = 4\pi\lambda_B(c_2 - c_1 - c_3) \tag{3.2}$$

where $s = 1$ for cylinder; $s = 2$ for sphere, and $\lambda_B$, is the Bjerrum length:

$$\lambda_B = \frac{e^2}{4\pi\varepsilon_0 \varepsilon k_B T} \tag{3.3}$$

Insofar as Eq. (3.2) is a second order differential equation, its general solution depends on two integration constants, $A_1$ and $A_2$:

$$\Psi = \Psi(r, A_1, A_2) \tag{3.4}$$

The following relations hold at the *outer* border of the cell:

$$\frac{\partial \Psi}{\partial r} = 0 \quad \text{and} \quad \Psi = 0 \quad \text{for } r = R_0 \tag{3.5a,b}$$

The first relation states that $\Psi$ has a local minimum at the border between two micelles; the second relation, $\Psi(R_0) = 0$, is based on the fact that the electric potential is defined up to an additive constant, which is set zero at the outer cell border. In addition, at the surface of micelle charges (of radius $R_{el}$) the following two relations take place:

$$\Psi_s = \Psi\big|_{r=R_{el}}, \quad \frac{\partial \Psi}{\partial r}\bigg|_{r=R_{el}} = -4\pi\lambda_B(\Gamma_1 - \Gamma_2) \tag{3.6a,b}$$

The first relation is the definition of the dimensionless surface potential $\Psi_s$. The second relation is the dimensionless form of the standard boundary condition relating the normal derivative of potential $\Psi$ with the surface charge density, which is proportional to $\Gamma_1 - \Gamma_2$. The counterion adsorption, $\Gamma_2$, is related to the subsurface activity of counterions, $a_{2s}$, by the Stern adsorption isotherm:

$$\frac{\Gamma_2}{\Gamma_1 - \Gamma_2} = K_{St}a_{2s} = K_{St}\gamma_{2,0}c_{2,0}\exp(\Psi_s) \tag{3.7}$$

Here, $c_{2,0}$ is the counterion concentration at the outer cell boundary ($r = R_0$); $\gamma_{2,0}$ is the respective activity coefficient; $K_{St}$ is the Stern constant, which can be determined from fits of surface tension isotherms or data for micelle aggregation number. The activity coefficient $\gamma_{2,0}$ is calculated as explained in Section 4.

At equilibrium, the electrochemical potentials are uniform throughout the EDL. In view of Eq. (2.18) and (3.1), this leads to a relation between the ionic concentrations in the EDL, $c_i = c_i(r)$, with their values at the outer cell border, $c_{i,0}$:

$$\ln(\gamma_i c_i) - (-1)^i \Psi = \ln(\gamma_{i,0} c_{i,0}) \quad (i = 1, 2, 3) \tag{3.8a,b,c}$$

where $\gamma_i = \gamma_i(c_1, c_2, c_3)$, $i = 1, 2, 3$, are local values of the activity coefficients in the EDL, which are calculated as described in Section 4. Correspondingly, $\gamma_{i,0} \equiv \gamma_i(c_{1,0}, c_{2,0}, c_{3,0})$, $i = 1, 2, 3$. As before, the subscripts 1, 2 and 3 number quantities, which are related, respectively, to surfactant ions, counterions and coions due to added salt. The subscript 0 denotes the values of the variables at the outer cell boundary, where $r = R_0$.

In the limiting case of diluted solutions, $\gamma_i = \gamma_{i,0} = 1$, Eqs. (3.8a,b,c) are reduced to the Boltzmann equations relating $c_i$ with $\Psi$. However, because the wormlike micelles from ionic surfactants grow in relatively concentrated electrolyte solutions, in general, we have to work with $\gamma_i \neq 1$ and $\gamma_{i,0} \neq 1$

Finally, to close the system of equations, we have to consider also the mass balances of surfactant and salt. The mass balance equations for surfactant ions, counterions and coions due to added salt read:

$$C_1 = (s+1)\frac{\Gamma_1}{R_{el}}(\frac{R_{el}}{R_0})^{(s+1)} + \frac{s+1}{R_0^{s+1}} \int_{R_{el}}^{R_0} c_1 r^s \, dr \tag{3.9}$$

$$C_2 = (s+1)\frac{\Gamma_2}{R_{el}}(\frac{R_{el}}{R_0})^{(s+1)} + \frac{s+1}{R_0^{s+1}} \int_{R_{el}}^{R_0} c_2 r^s \, dr \tag{3.10}$$

$$C_3 = \frac{s+1}{R_0^{s+1}} \int_{R_{el}}^{R_0} c_3 r^s \, dr \tag{3.11}$$

As usual, $s = 1$ for cylinder; $s = 2$ for sphere. The first terms in Eqs. (3.9) and (3.10) take into account contributions, respectively, from surfactant ions incorporated in the micelles and of counterions bound in the micelle Stern layer. The integral terms in the above three equations take into account contributions from the diffuse part of the EDL, which is located in the domain $R_{el} \leq r \leq R_0$. In Eq. (3.11), there is no "adsorption" term because binding of coions to the (like charged) surfactant headgroups is not expected.

In the case of spherocylindrical (wormlike) micelles, these mass balances have to be formulated for the *cylindrical parts* of the micelles ($s = 1$), insofar as we consider long micelles, for which the contribution of the endcaps to the total mass balance is negligible.

Note that Eqs. (3.9), (3.10) and (3.11) are not independent. Indeed, if these equations are substituted in the electroneutrality condition $C_2 - C_1 - C_3 = 0$, and $c_2 - c_1 - c_3$ is substituted from the Poisson equation, Eq. (3.2), one obtains Eq. (3.6b). Hence, only two among Eqs. (3.9), (3.10) and (3.11) are independent.

In the case of spherocylindrical micelles, the formulation of the electrostatic boundary-value problem is different for the cylindrical part and for the endcaps, as follows.

(A) *Cylindrical part (s = 1)*. At given $C_1$, $C_3$, $R_{el}$, $\Gamma_1$ and $c_{1,0}$, Eqs. (3.4), (3.5a,b), (3.6a,b), (3.7), (3.8a,b,c), (3.10), and (3.11) form a system of 11 equations for determining the following 11 unknowns: $\Psi$, $\Psi_s$, $A_1$, $A_2$, $\Gamma_2$, $R_0$, $c_1$, $c_2$, $c_3$, $c_{2,0}$, and $c_{3,0}$. The algorithm for solving this problem can be found in SI Appendix F. This procedure is applicable also to charged spherical micelles ($s = 2$).

(B) *Endcaps (s = 2)*. The concentrations at the outer border of the cell, $c_{2,0}$ and $c_{3,0}$, have been already determined from the solution of the problem for the cylindrical parts of the micelles (see above). In such a case, the input parameters are $C_1$, $C_3$, $R_{el}$, $\Gamma_1$, $c_{1,0}$, $c_{2,0}$ and $c_{3,0}$. Then, Eqs. (3.4), (3.5a,b), (3.6a,b), (3.7), (3.8a,b,c) form a system of 9 equations for determining of the following 9 unknowns: $\Psi$, $\Psi_s$, $A_1$, $A_2$, $\Gamma_2$, $R_0$, $c_1$, $c_2$, and $c_3$. The algorithm for solving this problem can be found in SI Appendix F.

(C) *Spherical micelles and CMC*. The cell model is applicable also to describe the micellar properties at the critical micellization concentration (CMC), at which the micelles are supposed to be spherical. At the CMC the concentration of micelles is low, so that we can set $R_0 \to \infty$. Then, instead of Eq. (3.5a,b), the following boundary condition takes place:

$$\Psi \to 0 \text{ for } r \to \infty \tag{3.12}$$

The input parameters are $c_{1,0}$, $c_{2,0}$, $c_{3,0}$, $\Gamma_1$, $R_{el}$, $\lambda_B$ and $K_{St}$. Eqs. (3.4), (3.6a,b), (3.7), (3.8a,b,c) and (3.12) form a system of 8 equations for determining of the following 8 unknowns: $\Psi$, $\Psi_s$, $A_1$, $A_2$, $\Gamma_2$, $c_1$, $c_2$, and $c_3$. The algorithm for solving this problem can be found in Ref. [4].

The solution of the electrostatic boundary-value problem for spherical micelles near the CMC has at least two applications. First, at given (experimental) CMC = $c_{1,0}$ without added salt ($c_{3,0} = 0$) the micellization energy $\Delta\mu_{mic}^o$ is determined. Second, at known $\Delta\mu_{mic}^o$, the dependence of the CMC on the concentration of added salt can be predicted. For details, see Ref. [4], Appendixes C and F therein.

## 4. Theoretical expressions for the activity coefficients of the ions

*4.1. Theoretical model*

Wormlike micelles from ionic surfactants are usually formed at high concentrations of added salt, which can be higher than 1 M. In addition, near the charged micelle surface the concentration of counterions can be considerably greater than their mean concentration. Under such conditions, the effect of interactions between the ions in the diffuse EDL and in the Stern layer must be taken into account. For this goal, the activity coefficients $\gamma_i$, which enter Eqs. (3.7) and (3.8), have to be calculated. Here, we work with individual activity coefficients for each kind of ions, which depend on the position in the EDL: $\gamma_i = \gamma_i(r)$.

At high concentrations, the distances between the ions are comparable with the ionic diameters. For this reason, the effect of the finite ionic size has to be taken into account. Insofar as different ions have different radii, the best way to quantify this effect is to use the theoretical expression for the activity coefficient of a mixture of hard spheres of different radii originating from the Boublik–Mansoori–Carnahan–Starling–Leland (BMCSL) equation of state [12,13]. In addition to the hard-sphere interactions, we have to take into account (i) the electrostatic interactions and (ii) the contribution of any other interactions, which will be termed "specific" interactions.

In the expression for the electrochemical potential, Eq. (2.17), the effects of the aforementioned interactions are incorporated in the term $k_B T \ln \gamma_i$. Correspondingly, this term can be presented as a sum of three contributions:

$$k_B T \ln \gamma_i = \mu_i^{(el)} + \mu_i^{(hs)} + \mu_i^{(sp)} \quad (i = 1, 2, 3) \tag{4.1}$$

Here, $\mu_i^{(el)}$ takes into account the electrostatic (Debye-Hückel type) interaction between the ions; $\mu_i^{(hs)}$ accounts for the hard-sphere interactions, and finally, $\mu_i^{(sp)}$ expresses the contribution of any other "specific" interactions. Then, the activity coefficient can be presented in the form:

$$\gamma_i = \gamma_i^{(el)} \gamma_i^{(hs)} \gamma_i^{(sp)} \quad (i = 1, 2, 3) \tag{4.2}$$

where the three multipliers correspond to the three additives in Eq. (4.1), e.g. $\mu_i^{(el)} = k_B T \ln \gamma_i^{(el)}$, etc.

As before, we use the convention that the subscripts 1, 2 and 3 denote quantities related, respectively, to the surfactant ions, counterions and coions due to the added salt. In fact, Eqs. (4.1) and (4.2), as well as Eqs. (4.3) and (4.6) below, are applicable to an arbitrary number of ionic components, $1 \le i \le n$, not necessarily $n = 3$.

To calculate $\gamma_i^{(el)}$, we used the expression [33]:

$$\ln \gamma_i^{(el)} = -\frac{z_i^2 \lambda_B}{b_i} [\frac{\kappa b_i - 2}{2\kappa b_i} + \frac{\ln(1+\kappa b_i)}{(\kappa b_i)^2}]$$

$$-2\pi \frac{z_i^2 \lambda_B^2}{\kappa} \sum_{j=1}^{3} \frac{z_j^2 c_j}{(\kappa b_j)^2} [\frac{2+\kappa b_j}{1+\kappa b_j} - \frac{2}{\kappa b_j} \ln(1+\kappa b_j)] \quad (i = 1, 2, 3) \tag{4.3}$$

where $b_i$ is the radius of the respective ion (close to its hydrated radius), $z_i$ is its valence, and $\kappa$ is the Debye parameter:

$$\kappa^2 = 4\pi \lambda_B \sum_{i=1}^{3} z_i^2 c_i \tag{4.4}$$

Note that in Eqs. (4.3) and (4.4), $c_i = c_i(r)$ are the local ionic concentrations in the EDL. Consequently, $\kappa = \kappa(r)$ and $\gamma_i^{(el)} = \gamma_i^{(el)}(r)$ also vary across the EDL. In other words, Eq. (4.3) generalizes the Debye-Hückel expression to the case of non-uniform solutions (like the EDL). If the ionic radii are equal, $b_1 = b_2 = \ldots = b_n = b$, Eq. (4.3) reduces to the simpler formula [33]:

$$\ln \gamma_i^{(el)} = -\frac{z_i^2 \kappa \lambda_B}{2(1+\kappa b)} \tag{4.5}$$

For a *uniform* solution, Eq. (4.5) coincides with the Debye-Hückel formula for the activity coefficient [32]. Eq. (4.5) can be used also in a *non-uniform* EDL with $\kappa = \kappa(r)$ and $\gamma_i^{(el)} = \gamma_i^{(el)}(r)$.

The finite ionic size can have a significant effect on the properties of the EDL, especially in the case of higher ionic strengths [48]. Here, to calculate $\gamma_i^{(\text{hs})}$ we will use the expression for the activity coefficient of a hard-sphere fluid composed of several components of different radii. This expression, which is derived from the BMCSL equation of state [12,13], reads [14]:

$$\ln \gamma_i^{(\text{hs})} = -\left(1 - 12r_i^2 \frac{\xi_2^2}{\xi_3^2} + 16r_i^3 \frac{\xi_2^3}{\xi_3^3}\right)\ln(1-\xi_3) + \frac{2r_i(3\xi_2 + 6r_i\xi_1 + 4r_i^2\xi_0)}{1-\xi_3}$$

$$+ 12r_i^2 \xi_2 \frac{\xi_2 + 2r_i\xi_1\xi_3}{\xi_3(1-\xi_3)^2} - 8r_i^3\xi_2^3 \frac{\xi_3^2 - 5\xi_3 + 2}{\xi_3^2(1-\xi_3)^3} \quad (i=1,\ 2,\ 3) \tag{4.6}$$

where

$$\xi_m \equiv \frac{\pi}{6}\sum_{i=1}^{3} c_i (2r_i)^m \quad (m=0,\ 1,\ 2,\ 3) \tag{4.7}$$

In Eq. (4.7), the index $i$ numbers the ionic components, whereas the index $m$ numbers the powers of the hard-sphere diameter, $(2r_i)$. For the ions in an electrolyte solution, in general, the radii $b_i$ in Eq. (4.3) and $r_i$ in Eq. (4.6) are different; see Table 1 below. In both Eq. (4.3) and Eq. (4.7), $c_i$ are number (rather than molar) concentrations.

Despite the high salt and surfactant concentrations in the micellar solutions, the water still has the highest molar fraction, at least ten times greater than that of the solutes. Then, we can expand in series the Wilson equation for mixed solutions (see Eq. (1.200) in Ref. [34], as well as Refs. [26] and [49]) in order to derive (in linear approximation) an expression for the specific interactions:

$$\ln \gamma_i^{(\text{sp})} = -2\sum_{j=1}^{3} \beta_{ij} c_j \quad (i=1,\ 2,\ 3) \tag{4.8}$$

Here, the summation is over the different kinds of ions in the solution and $\beta_{ij} = \beta_{ji}$ are interaction parameters.

To avoid using many adjustable parameters, we can further simplify Eq. (4.8). Insofar as the like-charged ions repel each other and are separated at greater distances, the predominant contribution to $\ln \gamma_i^{(\text{sp})}$ is expected to come from the oppositely charged ions, which can come into close contact. In addition, because in solutions with WLM the concentration of free surfactant ions is much lower than that of the coions due to salt, a reasonable approximation is $\beta_{12} \approx \beta_{32} \equiv \beta$. Then, Eq. (4.8) acquires the following simpler form:

$$\ln \gamma_1^{(sp)} \approx \ln \gamma_3^{(sp)} = -2\beta c_2, \quad \ln \gamma_2^{(sp)} \approx -2\beta(c_1 + c_3) \tag{4.9}$$

*4.2. Determination of the parameters of the model*

To determine the values of the ionic radii $b_i$ and $r_i$, and the interaction parameter $\beta$ for the most frequently used electrolytes, NaCl, NaBr, KBr and KCl, we fitted literature data for the respective mean activity coefficient $\gamma_\pm = (\gamma_2 \gamma_3)^{1/2}$ (only salt; no surfactant) by using Eqs. (4.3), (4.6) and (4.9). We used experimental data for the dependence of $\gamma_\pm$ on the ionic strength, $I$, from the book by Robinson and Stokes [35]. For the needs of the present theoretical study, the molality-scale activity coefficients tabulated in Ref. [35], have been converted into molarity-scale activity coefficients used here; see SI Appendix G.

Initially, we varied five adjustable parameters: $b_2$, $b_3$, $r_2$, $r_3$, and $\beta$. The results showed that for the best fit (i) $b_2 \approx b_3 \equiv b$, and (ii) the values of $r_2$ and $r_3$ are very close to the hard-sphere radii of the respective bare ions as given in Ref. [50]. The fact that $b_2 \approx b_3 \equiv b$ probably means that the main contribution to $\gamma_i^{(el)}$ comes from the close contacts in the cationic-anionic pairs, and then $2b$ can be interpreted as the distance between the centers of the ions in such pairs upon contact; see Refs. [32,33].

The above result allowed us to fix $r_2$ and $r_3$ equal to the hard-sphere radii of the bare ions in Ref. [50], and to fit the data for $\gamma_\pm$ by using only two adjustable parameters: $b$ and $\beta$. In particular, for $b_2 = b_3 \equiv b$ Eq. (4.3) reduces to the simpler Eq. (4.5). The fits of experimental data for NaCl, KCl, NaBr and KBr are shown in Fig. 2 and the values of $b$ and $\beta$ determined from the best fits are given in Table 1, together with the *hard*-sphere radii of the cations and anions, $r_+$ and $r_-$ from Ref. [50]. In general, one sees that $b > r_+ + r_-$. This means that the value of $b$ includes a contribution from the hydration water. In the framework of 6-7 %, the values of $b$ coincide with the sum of the *soft*-sphere radii of the respective cation and anion given in Ref [50].

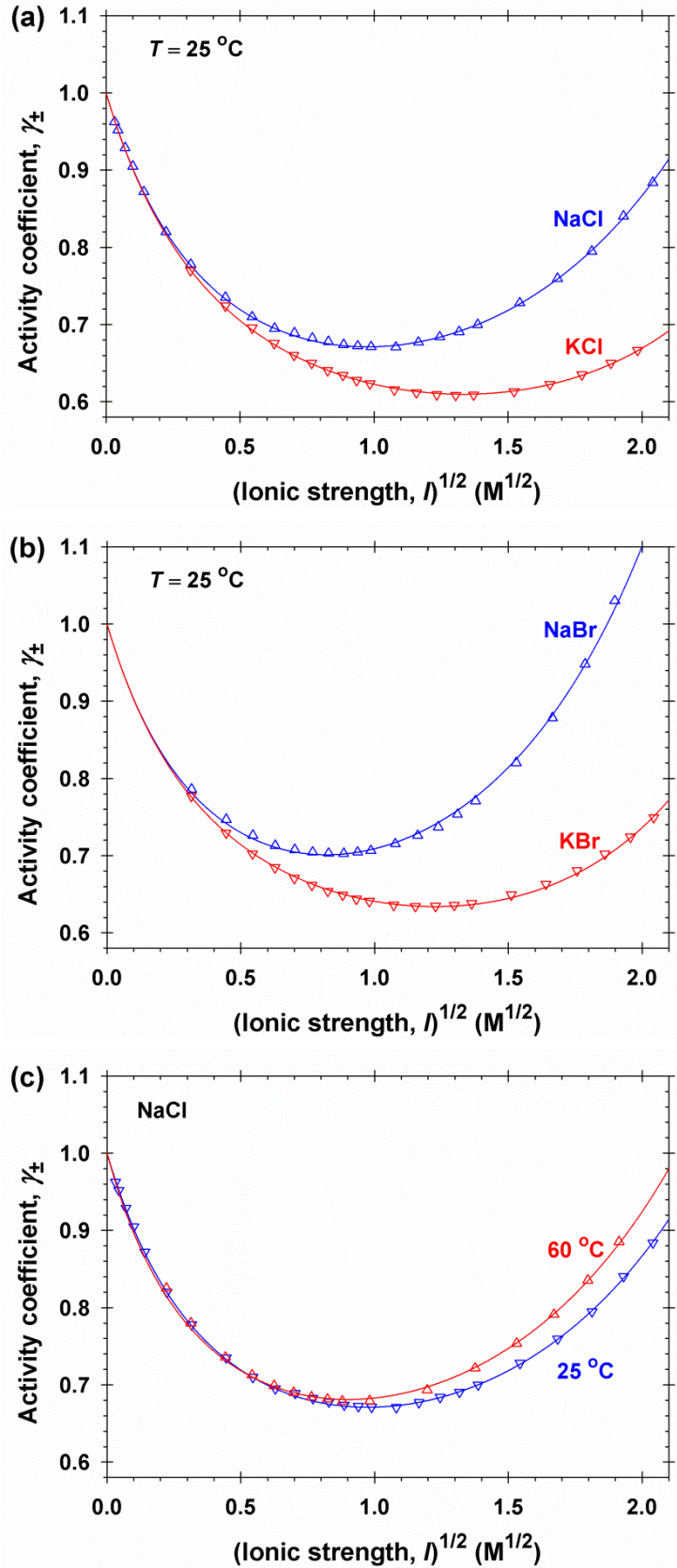

**Fig. 2.** Molarity-scale mean activity coefficient, $\gamma_\pm$, plotted vs. the square root of the ionic strength, $I^{1/2}$, for bulk electrolyte solutions. The points are experimental data from Ref. [35], whereas the solid lines are the best fits with the model in Section 4. (a) NaCl and KCl at 25 °C; (b) NaBr and KBr at 25 °C, and (c) NaCl at 25 and 60 °C.

**Table 1**. Parameters of the model used to calculate the activity coefficients $\gamma_i$ for $i = 2$ and 3: $b$ and $\beta$ are determined from the best fits of literature data [35] for $\gamma_\pm$ for alkali metal halides shown in the first column; $r_+$ and $r_-$ are literature data for the bare ionic radii [50].

| salt | $T$ (°C) | $b$ (Å) | $r_+$ (Å) | $r_-$ (Å) | $\beta$ (M$^{-1}$) |
|---|---|---|---|---|---|
| NaCl | 60° | 4.47 | 1.009 | 1.822 | 0.00466 |
| NaCl | 25° | 3.95 | 1.009 | 1.822 | 0.00966 |
| NaBr | 25° | 4.02 | 1.009 | 1.983 | $\approx 0$ |
| KCl | 25° | 3.90 | 1.320 | 1.822 | 0.0635 |
| KBr | 25° | 4.28 | 1.320 | 1.983 | 0.0789 |

In this paper, by definition $r_2$ and $r_3$ are the radii of the counterions and coions in a micellar solution, whereas in Table 1 $r_+$ and $r_-$ are radii of cations and anions. Thus, in the case of an *anionic* surfactant, e.g. sodium dodecyl sulfate (SDS) + NaCl, we have $r_2 = r_+$ and $r_3 = r_-$, where $r_+$ and $r_-$ are the values for NaCl in Table 1.

Fig. 2 shows the fits of experimental data, from which the values of $b$ and $\beta$ in Table 1 have been determined as adjustable parameters. As seen, the model excellently fits the experimental data. Figs. 2a and b show that $\gamma_\pm$ is markedly lower for the potassium salts as compared to the respective sodium salts. In the model, this difference is taken into account by the values of the interaction parameter $\beta$, which is significantly greater for the potassium salts (Table 1). Physically, this means that the specific interaction of the K$^+$ ions with the halide anions, Cl$^-$ and Br$^-$, is significantly stronger than that of the Na$^+$ ions.

In addition, Fig. 2c shows $\gamma_\pm$ for SDS at two different temperatures, 25 and 60 °C. One sees that the effect of temperature on $\gamma_\pm$ is not so significant, but it is not negligible because of the high sensitivity of the scission energy of the wormlike micelles to the thermodynamic state of the system; see Section 5. The data in Table 1 show that at 60 °C the parameter $b$ is greater, whereas $\beta$ is smaller, than its value at 25 °C. This difference could be explained with the stronger thermal motion at 60 °C, which leads to greater average separation between the cations and anions at this higher temperature.

In Fig. 3a, using KBr as an example, we compare the contributions of the different interactions in $\gamma_\pm$. The repulsive hard-sphere interactions lead to $\gamma_\pm^{(hs)} > 1$, whereas the attractive cation-anion electric and specific interactions lead to $\gamma_\pm^{(el)} < 1$ and $\gamma_\pm^{(sp)} < 1$. The non-monotonic dependence of $\gamma_\pm$ on $I$ is related to the significant rise of $\gamma_\pm^{(hs)}$ at higher ionic strengths. A numerical example: at 4 M KBr we have $\gamma_\pm^{(el)} = 0.5392$, $\gamma_\pm^{(hs)} = 2.567$, and $\gamma_\pm^{(sp)} = 0.5319$, so that $\gamma_\pm = \gamma_\pm^{(el)} \gamma_\pm^{(hs)} \gamma_\pm^{(sp)} = 0.7362$.

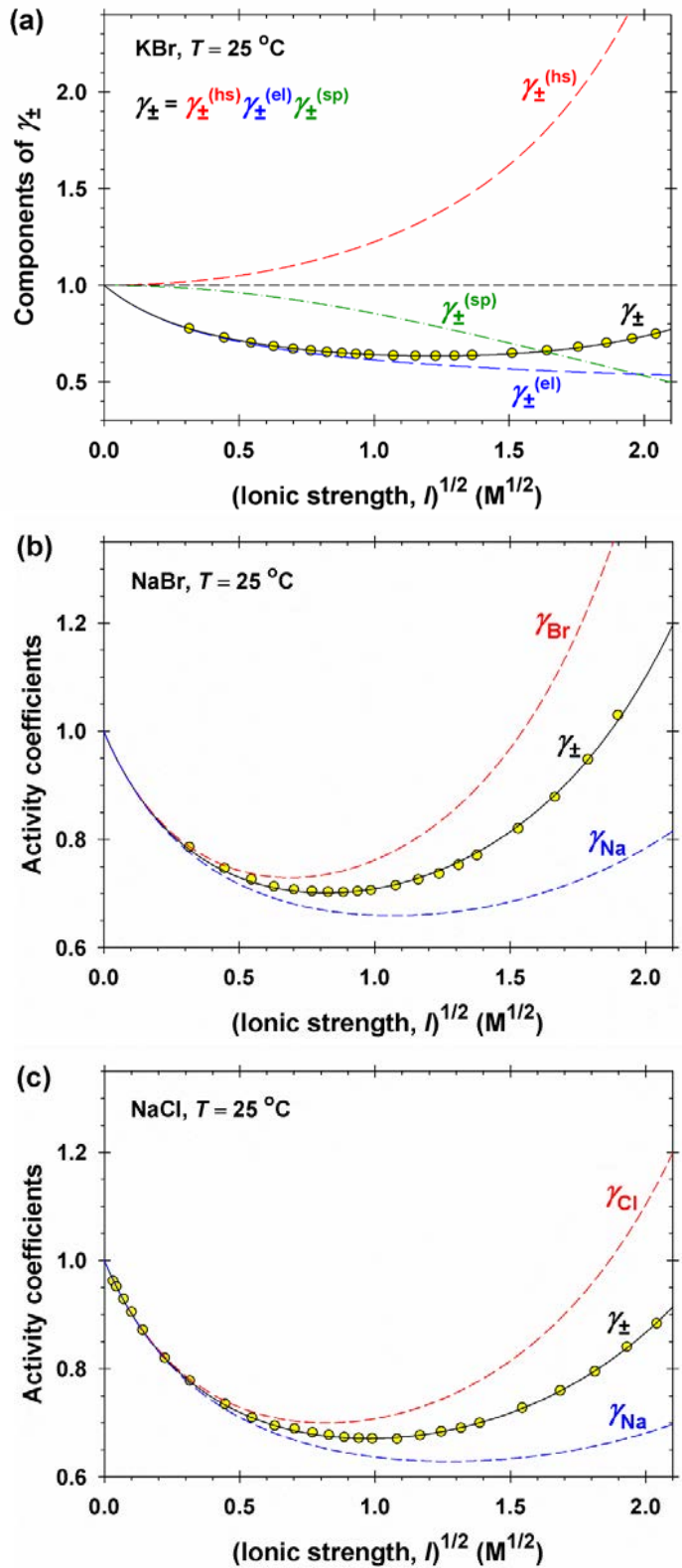

**Fig. 3.** Plots of molarity-scale mean activity coefficients vs. the square root of the ionic strength, $I^{1/2}$, for bulk electrolyte solutions at 25 °C. The points are experimental data [35]; the curves are calculated with the parameter values determined from the best fit (Table 1). (a) Comparison of the three components of $\gamma_\pm$, viz. $\gamma_\pm^{(el)}$, $\gamma_\pm^{(hs)}$ and $\gamma_\pm^{(sp)}$, for KBr. (b) Comparison of $\gamma_{Na}$, $\gamma_{Br}$ and $\gamma_\pm$ for NaBr. (c) Comparison of $\gamma_{Na}$, $\gamma_{Cl}$ and $\gamma_\pm$ for NaCl.

In Figs. 3b and c, the activity coefficients of the anions and cations are compared with the mean activity coefficient, $\gamma_\pm$, for NaBr and NaCl. One sees that for $I^{1/2} > 0.5$ M$^{1/2}$ (that is for $I > 0.25$ M – the range where WLM grow), there is significant difference between the activity coefficients of anions and cations. At that, $\gamma_{Br}$, $\gamma_{Cl} > \gamma_{Na}$, which is due to the greater size of the anions – see Table 1. The use of the correct values of the activity coefficients is a prerequisite for correct prediction of the electrostatic free energy of the wormlike micelle and its scission energy; see Section 5.

Note that the theoretical approach based on Eqs. (4.3), (4.6) and (4.8) allows one to predict the local activities of the various ions within the EDL, $\gamma_i = \gamma_i(r)$, whereas the semiempirical approach developed by Pitzer [26,31] predicts only the mean activity coefficient of uniform solutions, $\gamma_\pm$.

To calculate the activity coefficient, $\gamma_1$, of the free surfactant ions, which appear with a low concentration in the EDL (much lower than that of salt), a reasonable approximation is that they can be treated as the coions due to salt (see parameters in Table 1) with the only difference that the effective radius, $r_1$, of the surfactant ion is greater. In Section 5, numerical examples for SDS + NaCl are considered, where we have used $r_1 = 4.65$ Å estimated on the basis of molecular size considerations.

## 5. Numerical results and discussion

Here, we present illustrative numerical results (obtained by means of the developed model) for the effect of different factors on the properties of micelles from ionic surfactants. The studied properties related to the EDL around the micelle are (i) micelle surface potential, $\psi_s$; (ii) subsurface concentration of counterions, $c_{2s}$; (iii) occupancy of the Stern layer with bound counterions $\theta = \Gamma_2/\Gamma_1$ – a parameter that is related to the surface charge density; (iv) the outer radius of the counterion atmosphere in the cell model, $R_0$, and (v) the electrostatic free energy per molecule in the micelle, $f_{el}$.

Effects of the following factors have been investigated: (i) salt concentration, $C_3$; (ii) micelle geometry, sphere vs. cylinder, and (iii) activity coefficients, $\gamma_i$ ($i = 1, 2, 3$).

In order to compare the predictions of the theoretical model with experimental data for the mean mass aggregation number of wormlike micelles, $n_M$, we have to calculate the total interaction free energy per molecule in the micelle, $f_{int}$, which is a sum of four components corresponding to different kinds of interactions of a surfactant molecule in the micelle [1,7]:

$$f_{int} \equiv f_\sigma + f_{conf} + f_{hs} + f_{el} \tag{5.1}$$

where $f_\sigma$ is the interfacial-tension component; $f_{conf}$ is the chain-conformation component; $f_{hs}$ is the headgroup-steric component, and finally, $f_{el}$ is the electrostatic component; see Eq. (2.31). Hence, in addition to the electrostatic free energy, we have to accurately calculate the other three free-energy components. The values of $R_{el}$ for the cylindrical part of the micelle and its endcaps have to be found by minimization of $f_{int}$. (In the general case, $R_{el}$ is greater for the endcaps as compared to the cylindrical part.) This is done in [4], where the theory is compared with data for $n_M$ and excellent agreement is achieved.

In the present article, which is focused on the calculation of $f_{el}$, our goal is limited to demonstration of the effects of the aforementioned factors on micellar properties related to the EDL. For this goal, as an illustrative system we are using the anionic surfactant sodium dodecyl sulfate (SDS) in the presence of added NaCl. For this system, the values of the input parameters are estimated in SI Appendix F6. In these illustrative calculations, it has been assumed that the radius of micelle hydrophobic core is equal to the extended dodecyl chain of SDS. Then, one obtains $1/\Gamma_1 = 88.4$ Å$^2$ for spherical micelles; $1/\Gamma_1 = 49.7$ Å$^2$ for cylindrical micelles, and $R_{el} = 19.8$ Å for both spherical and cylindrical micelles. For the Stern constant, the value $K_{St} = 0.668$ M$^{-1}$ [51] was used. Insofar as the results are not sensitive to the concentration of free surfactant anions at the outer cell boundary (at $r = R_0$), in the present illustrative calculations we used a typical value, viz. $c_{1,0} = 5$ mM.

Fig. 4 shows the variation of the activity coefficients of the Na$^+$ and Cl$^-$ ions across the EDL of a cylindrical SDS micelle ($R_{el} \leq r \leq R_0$) at three NaCl concentrations, 0.5, 1.0 and 1.5 M. The ionic activity coefficients, $\gamma_i$, have been calculated by means of the full theory in Section 4, i.e., $\gamma_i = \gamma_i^{(el)} \gamma_i^{(hs)} \gamma_i^{(sp)}$, where the three components of $\gamma_i$ are calculated from Eqs. (4.5), (4.6) and (4.9) using the parameter values in Table 1. Greater deviation of $\gamma_{Na}$ and $\gamma_{Cl}$ from 1 indicate stronger effect of ionic interactions.

In general, the behavior of the dependences $\gamma_{Na}(r)$ and $\gamma_{Cl}(r)$ in the nonuniform EDL is rather different from that in a uniform solution – compare Fig. 3c with Fig. 4. Indeed, across the EDL the concentration of the Na$^+$ counterions increases monotonically up to $c_{2s} = 5.6$ M in the subsurface layer (Fig. 5b for 1.5 NaCl). However, both $\gamma_{Na}$ and $\gamma_{Cl}$ level off at greater distances from micelle surface and exhibit a pronounced variation near the charged micelle surface (Fig. 4). The latter variation is important, because it determines the subsurface activity of the counterions, $a_{2s} = \gamma_{2s} c_{2s}$, which (in turns) affects the occupancy of the Stern layer, $\theta = \Gamma_2/\Gamma_1$ (see Eq. (3.7)), and the net surface charge of the micelle.

Note also that in Fig. 4 the plateau values of $\gamma_{Na}$ and $\gamma_{Cl}$ are (in general) different from the bulk values in a uniform NaCl solution of the same concentration. Thus, at 0.5 M NaCl in the uniform solution we have $\gamma_{Na} \approx \gamma_{Cl} \approx 0.80$ (Fig. 3c), whereas in Fig. 4 the respective plateau values are $\gamma_{Na} = 0.67$ and $\gamma_{Cl} = 0.70$.

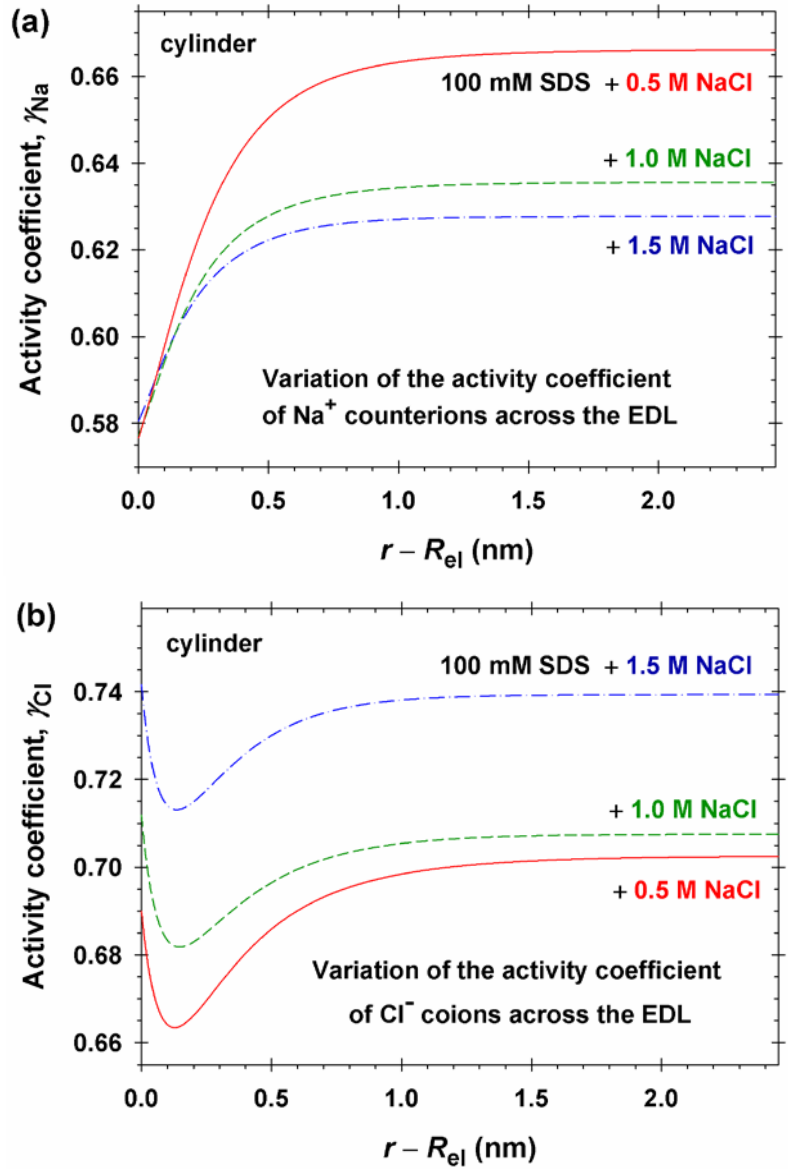

**Fig. 4.** Calculated variations of the activity coefficients of (a) the $Na^+$ counterions and (b) $Cl^-$ coions across the electric double layer ($R_{el} \leq r \leq R_0$) of the cylindrical micelles in 100 mM SDS solution at three NaCl concentrations, 0.5, 1.0 and 1.5 M; the right end of each plot corresponds to $r = R_0$.

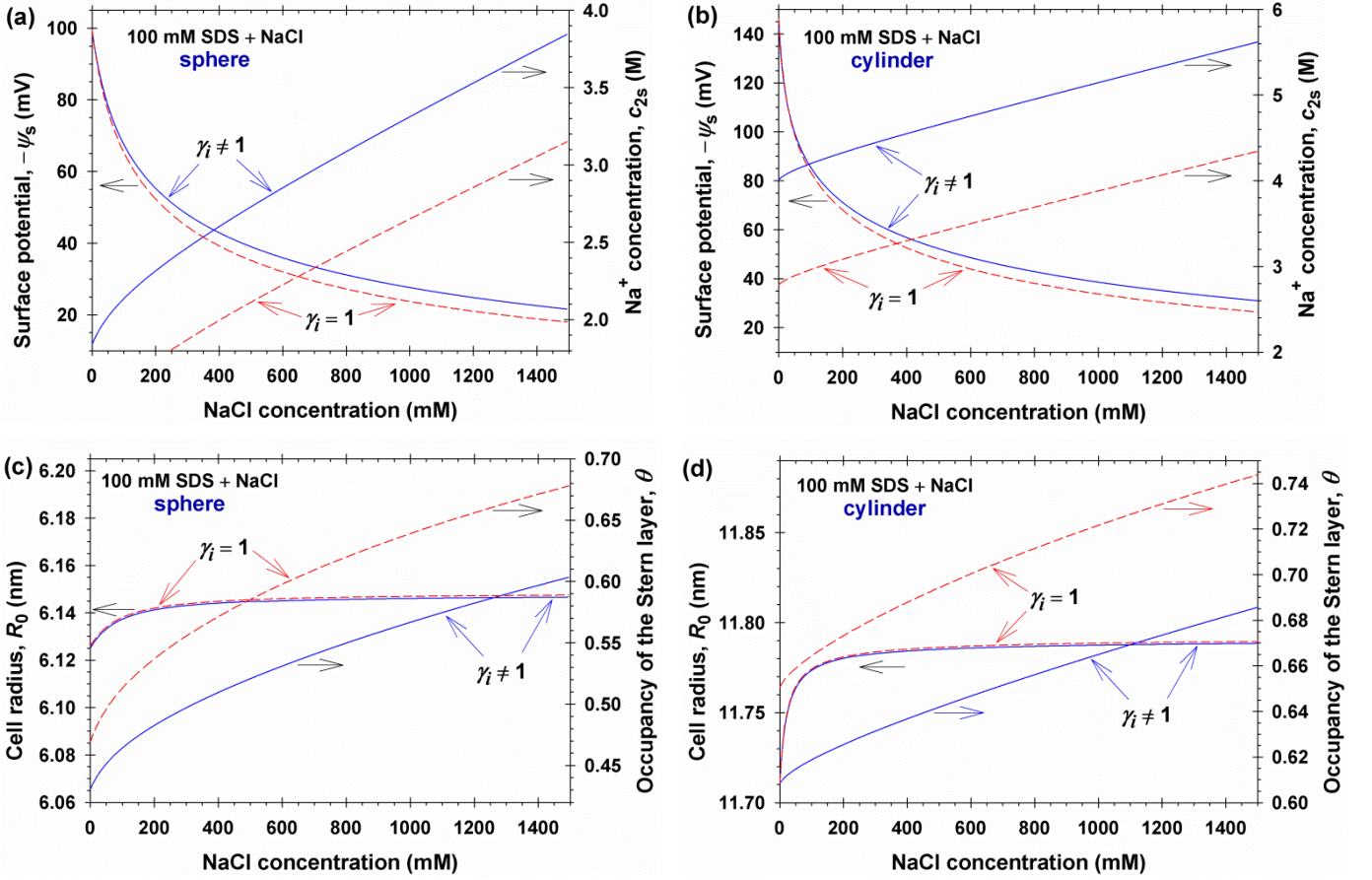

**Fig. 5.** Comparison of theoretical curves calculated taking into account the interactions between the ions in the EDL ($\gamma_i \neq 1$) with curves calculated neglecting these interactions ($\gamma_i = 1$) for micelles formed in 100 mM SDS solution with added NaCl. (a,b) Plots of the magnitude of the micelle surface potential, $-\psi_s$, and the subsurface $Na^+$ concentration, $c_{2s}$, vs. the NaCl concentration, $C_3$, for (a) spherical and (b) cylindrical micelle. (c,d) Plots of the cell radius, $R_0$, and the occupancy of the Stern layer, $\theta$, vs. $C_3$, for (c) spherical and (d) cylindrical micelle.

In Fig. 5, we compare theoretical curves calculated for $\gamma_i \neq 1$ and $\gamma_i = 1$. Here, $\gamma_i \neq 1$ means that the activity coefficients of the ions, $\gamma_i$, are calculated by means of the full theory in Section 4, as in Fig. 4. For the curves calculated with $\gamma_i = 1$ (shown with dashed lines), the interactions between the ions in the EDL have been neglected (ideal solution). All theoretical curves are calculated for the same surfactant concentration, $C_1 = 100$ mM SDS.

Figs. 5a and b illustrate the effect of NaCl concentration on the micelle surface electric potential, $\psi_s$, and on the subsurface concentration of $Na^+$ counterions, $c_{2s}$, for spherical and cylindrical micelles. As expected, $\psi_s$ decreases, whereas $c_{2s}$ increases with the rise of salt concentration. The values of both $\psi_s$ and $c_{2s}$ are higher for the cylindrical micelles as compared to the spherical ones. This is due to the higher density of charged surfactant headgroups, $\Gamma_1$, for the cylindrical micelles. The effect of activity coefficient, $\gamma_i$, is much stronger for $c_{2s}$ as compared to $\psi_s$. For cylindrical micelles, the subsurface Na$^+$ concentration, $c_{2s}$, is with up to 1.2 M higher for $\gamma_i \neq 1$, in comparison with the case of $\gamma_i = 1$. The highest computed value is $c_{2s} = 5.6$ M for cylindrical micelles at 1.5 M NaCl.

To check how important is the effect of counterion binding, we calculated also the subsurface Na$^+$ concentration assuming $\Gamma_2 = 0$ (no counterion binding); the results was $c_{2s} = 29$ M for cylindrical micelles at 0.5 M NaCl for the case with $\gamma_i = 1$. This completely non-physical result confirms the necessity to take into account the effect of counterion binding ($\Gamma_2 > 0$) and (ii) the effect of ionic interactions ($\gamma_i \neq 1$).

Figs. 5c and d illustrate the effect of NaCl concentration on the outer radius of the EDL, $R_0$, and on the occupancy of the Stern layer with bound Na$^+$ ions, $\theta = \Gamma_2/\Gamma_1$. One sees that (at fixed surfactant concentration) $R_0$ has a limited variation with the NaCl concentration, and levels off at $R_0 \approx 6.14$ nm for the spherical micelles and $R_0 \approx 11.77$ nm for the cylindrical ones. This behavior of $R_0$ can be understood by using the inequality (see SI Appendix F2):

$$\frac{(s+1)\Gamma_1}{R_{el}C_1} < \left(\frac{R_0}{R_{el}}\right)^{s+1} < \frac{(s+1)\Gamma_1 - R_{el}c_{1,0}}{R_{el}(C_1 - c_{1,0})} \tag{5.2}$$

which follows from the surfactant mass balance ($s = 1$ for cylindrical and $s = 2$ for spherical micelles). The relatively small value of $c_{1,0}$ leads to a relatively small range of variation of the calculated $R_0$. (For $c_{1,0} \to 0$, the two limits of $R_0/R_{el}$ coincide.) Then, the difference between the $R_0$ values for sphere and cylinder in Figs. 5c and d are related to the different values of $\Gamma_1$ and $s$ for spherical and cylindrical micelles. (Here, we work at fixed $R_{el} = 19.8$ Å and $c_{1,0} = 5$ mM.) Eq. (5.2) shows also that $R_0$ should decrease with the rise of the surfactant concentration $C_1$, which is related to the mutual confinement of the counterion atmospheres of the neighboring micelles in the solution (Fig. 1).

Figs. 5c and d show also that the effect of activity coefficient $\gamma_i$ on the occupancy of the Stern layer $\theta$ (and on the net surface charge) is significant: $\theta$ is with up to 6 – 7 % higher in the case $\gamma_i = 1$ as compared to $\gamma_i \neq 1$. This result might seem surprising in view of the opposite tendency for $c_{2s}$ in Figs. 5a and b. In fact, $\theta$ grows with the subsurface activity, $a_{2s} = \gamma_{2s}c_{2s}$, and it turns out that the effect of $\gamma_{2s}$ prevails – see the lower values of $\gamma_{Na} = \gamma_2$ near the micelle surface ($r - R_{el} = 0$) in Fig. 4a. Note also that $\theta$ essentially increases (the net surface charge density of the micelle, $z_1 e \Gamma_1(1-\theta)$, essentially decreases) with the rise of NaCl concentration.

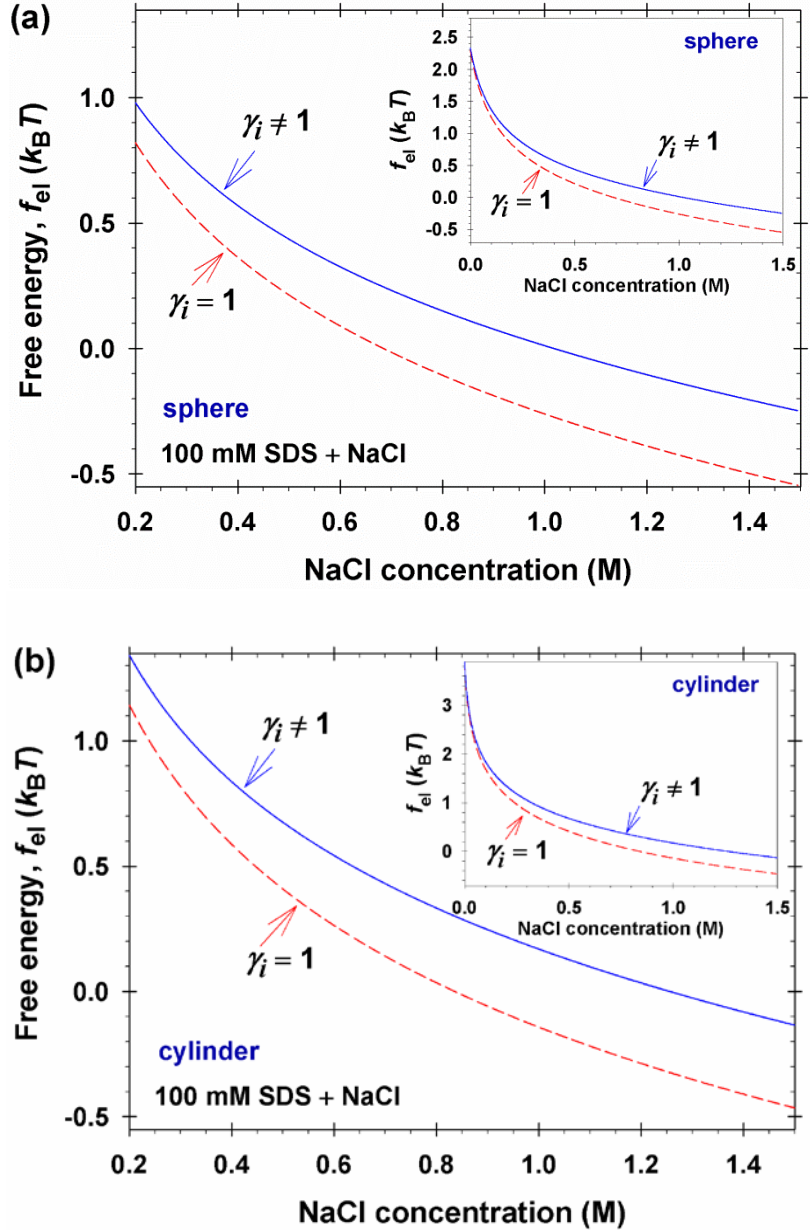

**Fig. 6.** Comparison of theoretical curves for $f_{el}$ vs. the NaCl concentration calculated taking into account the interactions between the ions in the EDL ($\gamma_i \neq 1$) with curves calculated neglecting these interactions ($\gamma_i = 1$); the micelles are formed in 100 mM SDS solution with added NaCl; $f_{el}$ is the electrostatic free energy per surfactant molecule in the micelle. (a) Spherical micelle. (b) Cylindrical micelle. The insets show the variation of $f_{el}$ in a wider range of NaCl concentrations.

Figs. 6a and b show plots of the electrostatic free energy per molecule, $f_{el}$, vs. the NaCl concentration, $C_3$, which are calculated using the same parameter values as in Fig. 5. As expected, $f_{el}$ decreases with the rise of $C_3$ because of the screening of the electrostatic interactions by the added electrolyte. At the highest salt concentrations, $f_{el}$ becomes negative, which is a consequence of the headgroup-counterion attraction in the Stern layer. The difference between the cases with $\gamma_i = 1$ and $\gamma_i \neq 1$ increases with the salt concentration and reaches ca. 0.3 $k_B T$ at 1.5 M NaCl. Is this difference physically important?

To answer this question, one could use an estimate based on the relation between the mean mass aggregation number of wormlike micelles, $n_M$, and the excess interaction free energy (scission energy) per molecule in the endcaps, $f_{sc}$ [7,52,53]:

$$n_M \approx 2(X_1 - X_1^o)^{1/2} \exp\left(\frac{n_s f_{sc}}{2k_B T}\right) \tag{5.3}$$

where $X_1$ is the total surfactant molar fraction in the solution; $X_1^o$ is the surfactant molar fraction at the CMC; $n_s$ is the aggregation number of the two micelle endcaps together. To estimate the error, $\Delta n_M$, of the aggregation number $n_M$, which is due to an error $\Delta f_{sc}$ in the value of $f_{sc}$, we differentiate Eq. (5.3):

$$\frac{\Delta n_M}{n_M} \approx \frac{n_s \Delta f_{sc}}{2k_B T} \tag{5.4}$$

With $n_s = 70$ and $\Delta f_{sc} = 0.3$ $k_B T$, Eq. (5.4) gives a relative error $\Delta n_M/n_M = 10.5$ (that is 1050 %). Using Eq. (5.4) and the same parameter values, one estimates that in order to determine the aggregation number $n_M$ with a relative error $\Delta n_M/n_M = 10$ %, the error in the value of $f_{sc}$ should be $\Delta f_{sc} = 0.003$ $k_B T$.

In view of the fact that $f_{el}$ is one of the components of $f_{sc}$, the above results clearly show why we have to determine $f_{el}$ with the maximal possible accuracy, and in particular, why the effect of activity coefficients $\gamma_i$ ($i = 1, 2, 3$) must be taken into account. The correct prediction of the $n_M$ values could seem a very difficult task, but as demonstrated in the next part of this study [4], this is achievable, even without using any adjustable parameters.

## 6. Conclusions

The goal of the present series of papers is to develop a molecular thermodynamic theory of the formation of wormlike micelles, which predicts their mean mass aggregation number in agreement with the experiment and gives quantitative description of the effect of all factors that influence the micellar growth. To achieve that, the theory has to calculate the excess free energy per surfactant molecule in the micelle endcaps (known also as scission energy) with high accuracy, better than 0.01 $k_BT$ (see Section 5), which is a considerable challenge.

The present article is devoted to the theory of growth of wormlike micelles from *ionic* surfactants in the presence of added salt. Here, we focus on the accurate calculation of micelle electrostatic free energy, $F_{el}$. The approximate assumption that micelle electrostatic potential decays at infinity, which has been used in previous studies [5-11] is removed. Instead, the electric field is calculated using a cell model, which takes into account the mutual spatial confinement of the EDLs of the neighboring micelles based on Poisson equation and the integral mass balances of surfactant and salt (Section 3). The effect of micelle surface curvature on the EDL is taken into account exactly, without using any truncated series expansions.

At high salt concentrations (0.4 – 4 M), at which WLMs form, the effect of activity coefficients $\gamma_i$ of the ions becomes important. In our study, theoretical expressions for $\gamma_i$ are used, which take into account (i) the electrostatic, (ii) the hard sphere, and (iii) the specific interactions between the ions, and exactly describe the concentration dependencies of the mean activity coefficients of electrolytes, $\gamma_\pm = (\gamma_+\gamma_-)^{1/2}$. A detailed model has been used, in which $\gamma_i$ varies across the EDL as a function of the local ionic concentrations. In addition, the effect of counterion binding has been taken into account via the Stern isotherm [36]. Such detailed description of the electrostatic effects with ionic surfactant micelles has been given in none of the preceding studies [5-11,15].

To take into account all aforementioned effects, we derived an appropriate expression for $F_{el}$ in terms of micelle electrostatic surface pressure, $\pi_{el}$; see Eq. (2.25). This expression, in combination with a new original computational procedure (SI Appendix F), allows one to quickly calculate the micelle electrostatic free energy with one-time solution of the boundary-value problem. The calculation of all theoretical curves reported in this paper is achievable with a standard laptop.

The presented numerical results (Section 5) illustrate the variation of quantities characterizing the EDL of cylindrical and spherical micelles with the rise of electrolyte (NaCl) concentration. The variation of the ionic activity coefficients, $\gamma_i$, across the EDL is also quantified (Fig. 4). The effect of $\gamma_i$ on the free energy per surfactant molecule in the micelle, $f_{el}$, leads to higher values of $f_{el}$ (as compared to the case with $\gamma_i = 1$, i.e., with neglected ionic interactions in the EDL). These results demonstrate that the effect of activity coefficients is essential for the correct prediction of the size of ionic wormlike micelles.

The obtained results are applied in the next paper of this series [4], where the present study on electrostatic effects is complemented with a molecular-thermodynamic study. The full micelle interaction free energy, Eq. (5.1), is minimized to obtain the equilibrium micelle shape; the results are compared with experimental data for the mean mass aggregation number of wormlike micelles from both anionic and cationic surfactants, and excellent agreement between theory and experiment is achieved. The perspective of this study is to extend it to mixed solutions of ionic, zwitterionic and nonionic surfactants in order to give a theoretical interpretation of the observed synergistic effects, which are manifested as peaks of viscosity.

**Acknowledgement**

The authors gratefully acknowledge the support from Unilever R&D, project No. MA-2018-00881N, and from the Operational Programme ''Science and Education for Smart Growth'', Bulgaria, project No. BG05M2OP001-1.001-0008.

**Supplementary Information**

Supplementary information associated with this article, containing Appendixes A – G, can be found, in the online version, at …

**List of Symbols**

Basic (or derived) SI units are given in parentheses for all quantities, except dimensionless ones.

| Symbol | Description |
|---|---|
| $a_{2s}$ (m$^{-3}$) | subsurface activity of the counterions |
| $A$ (m$^2$) | surface area |
| $A_1, A_2$ | constants of integration |
| $b_i$ (m) | hydrated (soft-sphere) radius of type $i$ ions |
| $c_i$ (m$^{-3}$) | local number concentration of the component $i$ |
| $c_{i,0}$ (m$^{-3}$) | the value of $c_i$ at $\psi = 0$ |
| $c_{2s}$ (m$^{-3}$) | subsurface concentration of counterions |
| $c_1$ (m$^{-3}$) | local number concentration of surfactant ions |
| $c_2$ (m$^{-3}$) | local number concentration of counterions |
| $c_3$ (m$^{-3}$) | local number concentration of coions due to salt |
| $C_1$ (m$^{-3}$) | total input number concentration of surfactant ions |
| $C_2$ (m$^{-3}$) | total input number concentration of counterions |
| $C_3$ (m$^{-3}$) | total input number concentration of coions due to salt |
| $e$ (C) | the magnitude of electronic charge |
| $\mathbf{E}$, $E$ (V m$^{-1}$) | vector of electric field and its magnitude |
| $E_k$ (V m$^{-1}$) | projection of $\mathbf{E}$ along the $k$-th coordinate axis |
| EDL | electric double layer |
| $f_b$ (J m$^{-3}$) | bulk density of the non-electrostatic contribution to EDL free energy |
| $f_{el}$ (J) | total electric free energy per surfactant molecule in the micelle |
| $\tilde{f}_{el}$ (J) | electric free energy of the diffuse EDL per surfactant molecule in the micelle |
| $f_{sc}$ (J) | excess interaction free energy (scission energy) per molecule in the endcaps of WLM |
| $F$ (J) | free energy |
| $F_{chem}$ (J) | chemical free energy component |
| $F_{el}$ (J) | electrostatic free energy component |
| $F_{mech}$ (J) | mechanical free energy component |
| $F_{EDL}$ (J) | total free energy of the electric double layer |
| $I$ (M) | ionic strength of solution |
| $k_B$ (J K$^{-1}$) | Boltzmann's constant |
| $K_{St}$ (m$^3$) | Stern constant |
| $n_M$ | mean mass aggregation number of surfactant micelles |
| $n_s$ | aggregation number of the two WLM endcaps together |
| $N_i$ | number of molecules from the component $i$ |
| $p$ (Pa) | isotropic hydrostatic pressure |
| $p_0$ (Pa) | the value of $p$ in the region with $\psi = 0$ |
| $P_{ik}$ (Pa) | Maxwell electric pressure tensor |
| $P_T$ (Pa) | tangential (with respect to the surface) component of $P_{ik}$ |
| $q$ (C) | electric charge of the surfactant ion |
| $q_i$ (C) | electric charge of the component $i$ |
| $r$ (m) | radial distance from the micelle center |
| $r_i$ (m) | hard-sphere radii of the bare type $i$ ions |
| $r_+$, $r_-$ (m) | hard-sphere radii of bare cations and anions |
| $R_0$ (m) | outer radius of the cell containing a micelle and its EDL |
| $R_{el}$ (m) | radius of the surface of charges of a micelle |
| $s$ | $s = 1$ for cylindrical and $s = 2$ for spherical geometry |
| $S$ (J K$^{-1}$) | entropy |
| $T$ (K) | thermodynamic temperature |
| $U_{el}$ (J) | electrostatic energy of the EDL |
| $V$ (m$^3$) | volume |
| WLM | wormlike micelle |
| $X_1$ | total surfactant molar fraction in the solution |
| $z_1$ | valence of surface ionized groups |
| $z_2$ | valence of the counterions |
| $\beta$ and $\beta_{ij}$ (m$^3$) | parameters of the specific ion-ion interaction |
| $\gamma_i$ | activity coefficient of the ionic component $i$ |
| $\gamma_{i,0}$ | the value of $\gamma_i$ at $\psi = 0$ |
| $\gamma_i^{(el)}$ | contribution of electrostatic interactions to $\gamma_i$ |
| $\gamma_i^{(hs)}$ | contribution of hard-sphere interactions to $\gamma_i$ |
| $\gamma_i^{(sp)}$ | contribution of specific interactions to $\gamma_i$ |
| $\gamma_+, \gamma_-$ | activity coefficients of cations and anions |
| $\gamma_\pm = (\gamma_+\gamma_-)^{1/2}$ | mean activity coefficient |
| $\Gamma_1$ (m$^{-2}$) | number of ionizable groups per unit area of micelle surface |
| $\Gamma_2$ (m$^{-2}$) | number of bound counterions per unit area of micelle surface |
| $\delta_{ik}$ | the Kronecker delta symbol |
| $\varepsilon$ | relative dielectric constant of the medium (water) |
| $\varepsilon_0$ (F m$^{-1}$) | electric permittivity of vacuum |
| $\theta = \Gamma_2/\Gamma_1$ | occupancy of the Stern layer; degree of counterion binding |
| $\kappa$ (m$^{-1}$) | reciprocal Debye length |
| $\lambda_B$ (m) | the Bjerrum length |
| $\mu_i$ (J) | chemical potential of the component $i$ |
| $\mu_i^{el}$ (J) | electrochemical potential of the component $i$ |
| $\pi_{el}$ | electrostatic surface pressure |
| $\rho_b$ (C m$^{-3}$) | bulk electric charge density |
| $\rho_s$ (C m$^{-2}$) | surface electric charge density |
| $\psi$ (V) | electrostatic potential |
| $\psi_s$ (V) | surface electrostatic potential |
| $\Psi = q\psi/(k_BT)$ | dimensionless electrostatic potential |
| $\Psi_s = q\psi_s/(k_BT)$ | dimensionless surface electrostatic potential |

# Supplementary Information

for the article

## Analytical modeling of micelle growth. 3. Electrostatic free energy of ionic wormlike micelles – effects of activity coefficients and spatially confined electric double layers

Authors: Krassimir D. Danov, Peter A. Kralchevsky, Simeon D. Stoyanov, Joanne L. Cook, and Ian P. Stott

Here, the reference numbers are different from those in the main text; the list of cited references is given at the end of the present Supplementary Material.

The equation numbers in Supplementary Information material begin with a capital letter, e.g. (A1), (A2), etc. Equations numbered (3.1), (3.2), etc. are equations from the main text of this article. WLM = wormlike micelle.

**Appendix A. Derivation of the two forms of Eq. (2.1)**

By definition, we have $\mathbf{E} = -\nabla \psi$. Then, $E^2 = (\nabla \psi)^2 = (\nabla \psi) \cdot (\nabla \psi)$. Differentiating, we get:

$$\nabla \cdot (\psi \nabla \psi) = (\nabla \psi) \cdot (\nabla \psi) + \psi \nabla^2 \psi \tag{A1}$$

In addition, the Poisson equation reads:

$$\varepsilon \varepsilon_0 \nabla^2 \psi = -\rho_b \tag{A2}$$

Then, using Eqs (A1) and (A2) we obtain [1]:

$$\begin{aligned}\frac{\varepsilon \varepsilon_0}{2} \int_V E^2 \, dV &= \frac{\varepsilon \varepsilon_0}{2} \int_V (\nabla \psi)^2 \, dV = \frac{\varepsilon \varepsilon_0}{2} \int_V [\nabla \cdot (\psi \nabla \psi) - \psi \nabla^2 \psi] \, dV \\ &= \frac{\varepsilon \varepsilon_0}{2} \int_A dA \, \mathbf{n} \cdot (\psi \nabla \psi) + \frac{1}{2} \int_V \psi \rho_b \, dV \\ &= \frac{1}{2} \int_A \rho_s \psi_s \, dA + \frac{1}{2} \int_V \rho_b \psi \, dV \end{aligned} \tag{A3}$$

Here, we have used the Gauss divergence theorem and the boundary condition

$$\varepsilon \varepsilon_0 \mathbf{n} \cdot \nabla \psi = \rho_s \quad \text{at} \quad \mathbf{r} \in A, \tag{A4}$$

where **n** is the outer unit normal to the surface $A$ of the volume $V$, and $\rho_s$ is the surface charge density.

**Appendix B. Derivation of Eq. (2.16) from Eq. (2.14)**

Eq. (2.14) reads:

$$F_{el} = \int_V [-\frac{\varepsilon_0 \varepsilon E^2}{2} + \int_0^\psi \rho_b(\tilde{\psi}) \, d\tilde{\psi}] \, dV + \int_A \rho_s \psi_s \, dA \tag{B1}$$

where $\tilde{\psi}$ is the electric potential in the role of integration variable. Using Eq. (A3), we obtain:

$$\int_A \rho_s \psi_s \, dA = \varepsilon_0 \varepsilon \int_V E^2 \, dV - \int_V \rho_b \psi \, dA$$
$$= \varepsilon_0 \varepsilon \int_V E^2 \, dV - \varepsilon_0 \varepsilon \int_V \psi \nabla \cdot \mathbf{E} \, dV \tag{B2}$$

where at the last step the Poisson equation, Eq. (A2) with $\mathbf{E} = -\nabla \psi$, has been used. Substitution of Eq. (B2) in Eq. (B1), along with $\rho_b = \varepsilon \varepsilon_0 \nabla \cdot \mathbf{E}$, yields Eq. (2.16):

$$F_{el} = \varepsilon \varepsilon_0 \int_V [\frac{E^2}{2} - \psi \nabla \cdot \mathbf{E} + \int_0^\psi (\nabla \cdot \mathbf{E}) \, d\tilde{\psi}] \, dV \tag{B3}$$

**Appendix C. Derivation of Eq. (2.19) from Eq. (2.16)**

Here, we consider a symmetrical system (sphere, cylinder, plane), for which the electric field and potential depend only on the magnitude of position vector, $r = |\mathbf{r}|$: $E = E(r)$, $\psi = \psi(r)$, and consequently, the function $E = E(\psi)$ is also defined. In Eq. (2.16), which is identical to Eq. (B3), we substitute the Poisson equation, $\varepsilon \varepsilon_0 \nabla \cdot \mathbf{E} = \rho_b$:

$$F_{el} = \varepsilon \varepsilon_0 \int_V [\varepsilon \varepsilon_0 \frac{E^2}{2} - \rho_b \psi + \int_0^\psi \rho_b(\tilde{\psi}) \, d\tilde{\psi}] \, dV$$
$$= \int_V [\varepsilon \varepsilon_0 \frac{E^2}{2} - \int_0^\psi d(\rho_b \tilde{\psi}) + \int_0^\psi \rho_b(\tilde{\psi}) \, d\tilde{\psi}] \, dV \tag{C1}$$
$$= \int_V [\varepsilon \varepsilon_0 \frac{E^2}{2} - \int_0^\psi \tilde{\psi} \frac{d\rho_b}{d\tilde{\psi}} \, d\tilde{\psi})] \, dV$$

Using again the Poisson equation, $\rho_b = \varepsilon \varepsilon_0 \nabla \cdot \mathbf{E}$, we get:

$$F_{el} = \varepsilon\varepsilon_0 \int_V \{\int_0^\psi [E\frac{dE}{d\tilde{\psi}} - \tilde{\psi}\frac{d(\nabla\cdot\mathbf{E})}{d\tilde{\psi}}]d\tilde{\psi}\}dV$$
$$= \varepsilon\varepsilon_0 \int_V \{\int_0^\psi [\nabla\tilde{\psi}\frac{d(\nabla\tilde{\psi})}{d\tilde{\psi}} + \tilde{\psi}\frac{d(\nabla^2\tilde{\psi})}{d\tilde{\psi}}]d\tilde{\psi}\}dV \quad (C2)$$

The "imaginary charging process" [1] is equivalent to replace the integration variable $\tilde{\psi}$ with a new integration variable, $\xi = \tilde{\psi}/\psi$, which varies between 0 and 1. In terms of the new variable, Eq. (C2) acquires the following form:

$$F_{el} = \varepsilon_0\varepsilon \int_0^1 \{\int_V [\nabla\tilde{\psi}\cdot\frac{\partial(\nabla\tilde{\psi})}{\partial\xi} + \tilde{\psi}\frac{\partial(\nabla^2\tilde{\psi})}{\partial\xi}]dV\}d\xi$$
$$= \varepsilon_0\varepsilon \int_0^1 \{\int_V [\nabla\tilde{\psi}\cdot\nabla(\frac{\partial\tilde{\psi}}{\partial\xi}) + \tilde{\psi}\nabla^2(\frac{\partial\tilde{\psi}}{\partial\xi})]dV\}d\xi$$
$$= \varepsilon_0\varepsilon \int_0^1 \{\int_V \nabla\cdot[\tilde{\psi}\nabla(\frac{\partial\tilde{\psi}}{\partial\xi})]dV\}d\xi$$
$$= \varepsilon_0\varepsilon \int_0^1 \{\int_A \mathbf{n}\cdot[\tilde{\psi}\nabla(\frac{\partial\tilde{\psi}}{\partial\xi})]dA\}d\xi \quad (C3)$$
$$= \varepsilon_0\varepsilon \int_0^1 \{\int_A [\psi_s \frac{\partial(\mathbf{n}\cdot\nabla\tilde{\psi})}{\partial\xi}]dA\}d\xi$$
$$= \int_0^1 [\int_A (\psi_s \frac{\partial\tilde{\rho}_s}{\partial\xi})dA]d\xi$$
$$= \int_A (\int_0^{\rho_s} \psi_s\, d\tilde{\rho}_s)dA$$

The last expression is Eq. (2.19) in the main text; $\tilde{\rho}_s$ is the surface charge density in the role of integration variable. We have used the boundary condition, Eq. (A4).

**Appendix D. Derivation of Eq. (2.28)**

The expression for the surface pressure, Eq. (2.24) in the main text, is:

$$\pi_{el} = \frac{1}{R_{el}^s} \int_{R_{el}}^{R_0} (\frac{\varepsilon\varepsilon_0}{2}E^2 + p - p_0)r^s dr \quad \text{(cylinder, sphere)} \quad (D1)$$

Here and hereafter, $s = 1$ for cylinder and $s = 2$ for sphere; $r$ is the radial distance. The Poisson equation reads:

$$\varepsilon\varepsilon_0 (\frac{d^2\psi}{dr^2} + \frac{s}{r}\frac{d\psi}{dr}) = -\rho_b \tag{D2}$$

Eq. (D2) is multiplied by $d\psi/dr$ and integrated from $r$ to $R_0$:

$$-\varepsilon\varepsilon_0 \frac{E^2}{2} + \varepsilon\varepsilon_0 \int_r^{R_0} \frac{s}{\tilde{r}} E^2 d\tilde{r} = -\int_0^{\psi} \rho_b \, d\tilde{\psi} = p_0 - p \tag{D3}$$

where Eq. (2.12) in the main text has been used. Next, $(p_0 - p)$ from Eq. (D3) is substituted in Eq. (D1):

$$\pi_{el} = \frac{\varepsilon\varepsilon_0}{R_{el}^s} \int_{R_{el}}^{R_0} (E^2 - \int_r^{R_0} \frac{s}{\tilde{r}} E^2 d\tilde{r}) r^s dr \quad \text{(cylinder, sphere)} \tag{D4}$$

The last term can be integrated by parts:

$$-\int_{R_{el}}^{R_0} (\int_r^{R_0} \frac{s}{\tilde{r}} E^2 d\tilde{r}) r^s dr = -\frac{s}{s+1} \int_{R_{el}}^{R_0} (\int_r^{R_0} \frac{E^2}{\tilde{r}} d\tilde{r}) dr^{s+1} = \frac{s}{s+1} \left( R_{el}^{s+1} \int_{R_{el}}^{R_0} \frac{E^2}{r} dr - \int_{R_{el}}^{R_0} E^2 r^s dr \right) \tag{D5}$$

Finally, the combination of Eqs. (D4) and (D5) yields

$$\pi_{el} = \varepsilon_0\varepsilon \int_{R_{el}}^{R_0} [\frac{1}{s+1}(\frac{r}{R_{el}})^s + \frac{s}{s+1}\frac{R_{el}}{r}] E^2 dr \tag{D6}$$

which is identical with Eq. (2.28) in the main text.

## Appendix E.
## Derivation of Eq. (2.28) from the general expression for surface pressure in Ref. [2]

In Ref. [2], by mechanical considerations in terms of the general pressure tensor, the following expression for the surface pressure has been derived for a spherical interface ($s = 2$)

$$\pi = \frac{1}{3} \int_{R_{el}}^{R_0} (P_T - P_N)[\frac{2R_{el}}{r} + (\frac{r}{R_{el}})^2] dr \tag{E1}$$

(we have used the fact that by definition the surface pressure equals the surface tension with the inverse sign); $P_T$ and $P_N$ are the tangential and normal components of the surface pressure tensor with respect to the interface; see Eq. (40) in Ref. [2].

The Maxwell electric pressure tensor is given by the expression [3]:

$$P_{ik} = (p + \frac{\varepsilon\varepsilon_0}{2} E^2)\delta_{ik} - \varepsilon\varepsilon_0 E_i E_k \quad (i,k = 1,2,3) \tag{E2}$$

where $\delta_{ik}$ is the Kronecker delta symbol (the unit matrix) and $E_i$ is the $i$-th component of the electric field **E**; $p$ is the local hydrostatic pressure. In the considered case of symmetric system, **E** is directed normal to the charged surface, so that

$$P_\text{T} = p + \frac{\varepsilon\varepsilon_0}{2}E^2, \quad P_\text{N} = p - \frac{\varepsilon\varepsilon_0}{2}E^2 \tag{E3}$$

Substituting Eq. (E3) in Eq. (E1), we obtain Eq. (D6) for $s = 2$ (spherical interface).

# Appendix F.
## General computation procedure for the electrostatic boundary-value problem

*F1. Input parameters*

We consider a solution of an ionic surfactant with added salt, where spherical or cylindrical (wormlike) micelles are formed. For the endcaps of the cylindrical micelles, which have the shape of truncated spheres, the electric field is calculated in spherical geometry (as for full spheres), i.e., the edge effects truncated-sphere/cylinder are neglected. This approximation is reasonable, as confirmed by the agreement theory/experiment achieved in the next part of this study, Ref. [4].

The input parameters are as follows:

$C_1$ – total surfactant concentration;
$C_3$ – total concentration of salt; the total concentration of counterions is $C_2 = C_1 + C_3$;
$R_\text{el}$ – radius of the surface, at which the micelle surface charges are located;
$\Gamma_1$ – surface density of surfactant charged headgroups at $r = R_\text{el}$;
$c_{1,0}$ – concentration of surfactant ions at $r = R_0$, i.e. at the outer boundary of the cell, which contains the EDL around the micelle; see Fig. 1 in the main text.
$s = 1$ for cylindrical micelles; $s = 2$ for spherical micelles and for the endcaps of WLMs.

Note: In the next part of this study [4], a procedure based on free-energy minimization is developed, which yields the values of $R_\text{el}$, $\Gamma_1$ and $c_{1,0}$ for each specific system.

The Stern constant, $K_\text{St}$, could be determined from experimental surface tension isotherms, or it could be found by fits of experimental data for the scission energy of wormlike micelles, $E_\text{sc}$; see Ref. [4].

*F2. Basic equations*

To solve numerically the Poisson equation in the cell model, it is convenient to introduce the dimensionless coordinate, $t$, as follows:

$$r \equiv R_0 - (R_0 - R_{el})t \tag{F1}$$

$t = 0$ corresponds to the outer cell boundary $r = R_0$, whereas $t = 1$ corresponds to the micelle surface, $r = R_{el}$.

In terms of the new variable $t$, the Poisson equation, Eq. (3.2), acquires the form:

$$\frac{d^2\Psi}{dt^2} - \frac{s(R_0 - R_{el})}{R_0 - (R_0 - R_{el})t}\frac{d\Psi}{dt} = 4\pi\lambda_B(R_0 - R_{el})^2(c_2 - c_1 - c_3) \quad \text{for } 0 < t < 1 \tag{F2}$$

The respective form of the boundary condition, Eq. (5.13), reads:

$$\frac{d\Psi}{dt} = 4\pi\lambda_B(R_0 - R_{el})\Gamma_1(1 - \frac{\Gamma_2}{\Gamma_1}) \quad \text{for } t = 1 \tag{F3}$$

where $\Gamma_2/\Gamma_1$ is calculated from the Stern isotherm for counterion adsorption, Eq. (3.7):

$$\frac{\Gamma_2}{\Gamma_1 - \Gamma_2} = K_{St}\gamma_{2,0}c_{2,0}\exp y_1(1) \tag{F4}$$

where $y_1(1) = \Psi_s$; see Eq. (F11) below. The mass balances for the surfactant ions and for the coions (due to the added salt) acquire the form:

$$C_1 = (s+1)\frac{\Gamma_1}{R_{el}}(\frac{R_{el}}{R_0})^{(s+1)} + (s+1)(1 - \frac{R_{el}}{R_0})^{s+1}\int_0^1 c_1(\frac{R_0}{R_0 - R_{el}} - t)^s \, dt \tag{F5}$$

$$C_3 = (s+1)(1 - \frac{R_{el}}{R_0})^{s+1}\int_0^1 c_3(\frac{R_0}{R_0 - R_{el}} - t)^s \, dt \tag{F6}$$

see Eqs. (3.9) and (3.11). Finally, the expression for the electrostatic component of micellar surface pressure reads:

$$\pi_{el} = \frac{k_BT}{4\pi\lambda_B(R_0 - R_{el})}\int_0^1\{\frac{1}{s+1}[\frac{R_0}{R_{el}} - (\frac{R_0}{R_{el}} - 1)t]^s + \frac{s}{s+1}[\frac{R_0}{R_{el}} - (\frac{R_0}{R_{el}} - 1)t]^{-1}\}(\frac{\partial\Psi}{\partial t})^2 dt \tag{F7}$$

see Eq. (2.28).

To determine the interval of variation of the cell radius $R_0$, we will use the inequality $0 < c_1 \leq c_{1,0}$ in combination with the surfactant mass balance, Eq. (F5), in the form

$$C_1 = (s+1)\frac{\Gamma_1}{R_{el}}(\frac{R_{el}}{R_0})^{(s+1)} + \frac{s+1}{R_0^{s+1}}\int_{R_{el}}^{R_0} c_1 r^s \, dr \tag{F8}$$

The resulting inequality reads:

$$(s+1)\frac{\Gamma_1}{R_{el}}(\frac{R_{el}}{R_0})^{(s+1)} < C_1 < (s+1)\frac{\Gamma_1}{R_{el}}(\frac{R_{el}}{R_0})^{(s+1)} + c_{1,0}(1 - \frac{R_{el}^{s+1}}{R_0^{s+1}}) \tag{F9}$$

Solving this inequality with respect to $R_0/R_{el}$, we obtain:

$$\frac{(s+1)\Gamma_1}{R_{el}C_1} < \left(\frac{R_0}{R_{el}}\right)^{s+1} < \frac{(s+1)\Gamma_1 - R_{el}c_{1,0}}{R_{el}(C_1 - c_{1,0})} \tag{F10}$$

*F3. Main computational module*

To calculate all parameters with a self-consistent precision, we define the following boundary-value (Cauchy) problem. The input parameters are $s$, $R_0/R_{el}$, $c_{1,0}$, $c_{2,0}$ and $c_{3,0}$.

(i) The functions $y_1(t)$ and $y_2(t)$ are defined as follows:

$$y_1(t) \equiv \Psi, \quad y_2(t) \equiv \frac{d\Psi}{dt} \tag{F11}$$

(ii) The functions $y_3(t)$ and $y_4(t)$ are related to the mass balances, Eqs. (F5) and (F6):

$$y_3(t) \equiv \int_0^t c_1(\tilde{t}) \left(\frac{R_0}{R_0 - R_{el}} - \tilde{t}\right)^s d\tilde{t}, \quad y_4(t) \equiv \int_0^t c_3(\tilde{t}) \left(\frac{R_0}{R_0 - R_{el}} - \tilde{t}\right)^s d\tilde{t} \tag{F12}$$

(iii) The function $y_5(t)$ is related to the mass balance, Eq. (3.10):

$$y_5(t) \equiv \int_0^t c_2(\tilde{t}) \left(\frac{R_0}{R_0 - R_{el}} - \tilde{t}\right)^s d\tilde{t} \tag{F13}$$

(iv) The function $y_6(t)$ is related to the calculation of $\pi_{el}$ in Eq. (F7):

$$y_6(t) \equiv \int_0^t \left\{\frac{1}{s+1}\left[\frac{R_0}{R_{el}} - \left(\frac{R_0}{R_{el}} - 1\right)\tilde{t}\right]^s + \frac{s}{s+1}\left[\frac{R_0}{R_{el}} - \left(\frac{R_0}{R_{el}} - 1\right)\tilde{t}\right]^{-1}\right\}\left(\frac{\partial\Psi}{\partial\tilde{t}}\right)^2 d\tilde{t} \tag{F14}$$

Hence, the numerical boundary-value problem reads:

$$\frac{dy_1}{dt} = y_2 \tag{F15}$$

$$\frac{dy_2}{dt} = \frac{s(R_0 - R_{el})}{R_0 - (R_0 - R_{el})t} y_2 + 4\pi\lambda_B (R_0 - R_{el})^2 (c_2 - c_1 - c_3) \tag{F16}$$

$$\frac{dy_3}{dt} = \left(\frac{R_0}{R_0 - R_{el}} - t\right)^s c_1 \tag{F17}$$

$$\frac{dy_4}{dt} = \left(\frac{R_0}{R_0 - R_{el}} - t\right)^s c_3 \tag{F18}$$

$$\frac{dy_5}{dt} = \left(\frac{R_0}{R_0 - R_{el}} - t\right)^s c_2 \tag{F19}$$

$$\frac{dy_6}{dt} = \left\{\frac{1}{s+1}\left[\frac{R_0}{R_{el}} - \left(\frac{R_0}{R_{el}} - 1\right)t\right]^s + \frac{s}{s+1}\left[\frac{R_0}{R_{el}} - \left(\frac{R_0}{R_{el}} - 1\right)t\right]^{-1}\right\} y_2^2 \tag{F20}$$

with simple boundary conditions:

$$y_1(0) = y_2(0) = y_3(0) = y_4(0) = y_5(0) = y_6(0) = 0 \tag{F21}$$

The boundary-value problem, Eqs. (F11)–(F21), is solved numerically using the Verner sixth-order coefficients for the Runge-Kutta method [5]. At each step of the numerical integration, the values of $c_1(t)$, $c_2(t)$ and $c_3(t)$ are determined by numerical solution of the equations:

$$\ln(\gamma_1 c_1) = \ln(\gamma_{1,0} c_{1,0}) - y_1 \,, \quad \ln(\gamma_2 c_2) = \ln(\gamma_{2,0} c_{2,0}) + y_1 \,, \quad \ln(\gamma_3 c_3) = \ln(\gamma_{3,0} c_{3,0}) - y_1 \qquad \text{(F22a)}$$

where the activity coefficients, $\gamma_i = \gamma_i(c_1,c_2,c_3)$, $i = 1,2,3$, are determined by Eqs. (4.2), (4.5), (4.6) and (4.9) with parameter values given in Table 1; see the main text.

In the special case $\gamma_i \equiv 1$, Eqs. (F22a) are transformed into explicit expressions for $c_1(t)$, $c_2(t)$ and $c_3(t)$ (Boltzmann equations):

$$c_1 = c_{1,0} \exp(-y_1) \,, \quad c_2 = c_{2,0} \exp(y_1) \,, \quad c_3 = c_{3,0} \exp(-y_1) \qquad \text{(F22b)}$$

*F4. Determination of $R_0$*

In view of Eq. (F11), the combination of Eqs. (F3) and (F4) yields:

$$y_2(1) = \frac{4\pi \lambda_B (R_0 - R_{el}) \Gamma_1}{1 + K_{St} \gamma_{2,0} c_{2,0} \exp(y_1(1))} \qquad \text{(F23)}$$

At given $c_{2,0}$ and $c_{3,0}$, Eq. (F23) is solved numerically (say by the bisection method), to determine $R_0/R_{el}$, which belongs to the interval in Eq. (F10). At each step of the numerical procedure, $y_1(1)$ and $y_2(1)$ are determined by running the module in Section F3 above.

*F5. Determination of $c_{2,0}$ and $c_{3,0}$*

To calculate $c_{2,0}$ and $c_{3,0}$, we use the mass balances for the counterions and coions, Eqs. (F6) and (3.10), which in view of Eqs. (F12) and (F13) can be presented in the form:

$$C_1 + C_3 = (s+1)\frac{\Gamma_2}{R_{el}}(\frac{R_{el}}{R_0})^{(s+1)} + (s+1)(1 - \frac{R_{el}}{R_0})^{s+1} y_5(1) \qquad \text{(F24)}$$

$$C_3 = (s+1)(1 - \frac{R_{el}}{R_0})^{s+1} y_4(1) \qquad \text{(F25)}$$

The values of $c_{2,0}$ and $c_{3,0}$ are obtained by numerical solution of Eqs. (F24) and (F25). This can be achieved, for example, by numerical minimization of a merit function, based on Eqs. (F24) and (F25). At each steps of the numerical procedure, we run the modules in Sections F3 and F4 to determine $R_0/R_{el}$, $y_4(1)$, $y_5(1)$, as well as $y_1(1)$ which enters the expression for $\Gamma_2$ in Eq. (F4).

Having determined $c_{2,0}$ and $c_{3,0}$, we calculate $\pi_{el}$:

$$\pi_{el} = \frac{k_B T}{4\pi\lambda_B (R_0 - R_{el})} y_6(1) \tag{F26}$$

see Eqs. (F7) and (F14). Finally, using Eqs. (2.31), (F4) and (F26) we calculate the free energy per ionic surfactant molecule in the micelle:

$$f_{el} = k_B T[y_1(1) + \ln(1 - \Gamma_2/\Gamma_1)] - \pi_{el}/\Gamma_1 \tag{F27}$$

*F6. Parameter values for the system SDS + NaCl at 25 °C*

At 25 °C, the length and volume of the dodecyl chain of SDS can be estimated from the Tanford formulas [6-8]:

$$l(n_C) = 2.8 + 1.265(n_C - 1) \text{ Å} \tag{F28}$$

$$v(n_C) = 54.3 + (n_C - 1)26.9 \text{ Å}^3 \tag{F29}$$

where $n_C$ is the number of C atoms in the paraffin tail. For $n_C = 12$, we get $l = 16.7$ Å and $v = 350$ Å$^3$. Because the radius of the sulfate headgroup is ca. 3.1 Å, the radius of the surface of charges can be estimated as $R_{el} = 16.7 + 3.1 = 19.8$ Å, supposedly, the micelle radius corresponds to extended paraffin chain. This value of $R_{el}$ will be used for the illustrative calculations in the present article. (In the next paper of this series [4], the equilibrium values of $R_{el}$, which are different for the cylindrical part of the WLM and its endcaps, are found by free-energy minimization).

For *spherical* micelles, the volume of the micellar core, $V_m$, and the micelle aggregation number, $N_{agg}$, are:

$$V_m = \frac{4}{3}\pi l^3, \quad N_{agg} = \frac{V_m}{v} = \frac{4\pi l^3}{3v} \tag{F30}$$

Then the density of surfactant headgroups on the surface of the micelle is:

$$\Gamma_1 = \left(\frac{4\pi R_{el}^2}{N_{agg}}\right)^{-1} = \frac{l^3}{3vR_{el}^2} \tag{F31}$$

With the above parameter values, we obtain $1/\Gamma_1 = 88.4$ Å$^2$ or $\Gamma_1 = 1.88 \times 10^{-6}$ mol/m$^2$.

In the case of cylindrical micelle of length $L$, we have:

$$V_m = \pi l^2 L, \quad N_{agg} = \frac{V_m}{v} = \frac{\pi l^2 L}{v} \tag{F32}$$

$$\Gamma_1 = \left(\frac{2\pi R_{el} L}{N_{agg}}\right)^{-1} = \frac{l^2}{2v R_{el}} \quad (F33)$$

With the above parameter values, we obtain $1/\Gamma_1$ = 49.7 Å$^2$ or $\Gamma_1$ = 3.34×10$^{-6}$ mol/m$^2$. As expected, $\Gamma_1$ is greater for the cylindrical micelle.

For SDS at 25 °C, we used the value of the Stern constant $K_{St}$ = 0.668 M$^{-1}$ ; see Ref. [9].

**Appendix G. Relation between the molality- and molarity-scale activity coefficients**

In the literature [10], molality-scale activity coefficients, $\gamma_m$, are given. In the present study, molarity-scale activity coefficients, $\gamma_\pm$, are used. For 1:1 electrolytes, the ionic strength $I$ (mol/l) and $\gamma_\pm$ are simply related to the molality $m$ (mol/kg) and $\gamma_m$ [10]:

$$I = \frac{m\rho}{1+mM/1000} \quad \text{and} \quad \gamma_\pm = \frac{m}{I}\gamma_m \quad (G1)$$

where $\rho$ (g/cm$^3$) is the density of the aqueous solution and $M$ (g/mol) is the molecular weight (Table G1).

**Table G1**. Molecular weights of alkali metal halides.

|  | NaCl | NaBr | KCl | KBr |
|---|---|---|---|---|
| $M$ (g/mol) | 58.443 | 102.894 | 74.551 | 119.002 |

We interpolated the experimental data for the density of NaCl aqueous solutions measured at 25 °C and 60 °C (Fig. G1a, symbols) and obtained the following interpolation formulae:

$$\rho = 0.99705 + 0.040228m - 1.3094\times10^{-3}m^2 \quad \text{at } 25\ °C \quad (G2)$$

$$\rho = 0.98320 + 0.038590m - 1.1814\times10^{-3}m^2 \quad \text{at } 60\ °C \quad (G3)$$

where the density is measured in g/cm$^3$ and the molality in mol/kg. The relative errors of predicted values are less than 2×10$^{-4}$ (see Fig. G1a – the solid lines).

The experimental data for the density, $\rho$ (g/cm$^3$), of NaBr aqueous solutions [11] versus the molality, $m$ (mol/kg), are interpolated as follows:

$$\rho = 0.99725 + 0.077531m - 2.2538\times10^{-3}m^2 \quad \text{at } 25\ ^\circ\text{C} \tag{G4}$$

The relative errors of predicted values for $m \leq 4$ mol/kg are less than $1.5\times10^{-4}$ (see Fig. G1b). The respective interpolation formula for the experimental data for KBr solutions [12] reads:

$$\rho = 0.99717 + 0.083812m - 3.4294\times10^{-3}m^2 \quad \text{at } 25\ ^\circ\text{C} \tag{G5}$$

The relative errors of predicted values for $m \leq 2.5$ mol/kg are less than $1.0\times10^{-4}$ (see Fig. G1b). Finally, the data for KCl solutions [13] are interpolated by the formula:

$$\rho = 0.99732 + 0.045428m - 1.6747\times10^{-3}m^2 \quad \text{at } 25\ ^\circ\text{C} \tag{G6}$$

The relative errors of predicted values for $m \leq 4.5$ mol/kg are less than $1.5\times10^{-4}$ (see Fig. C1b).

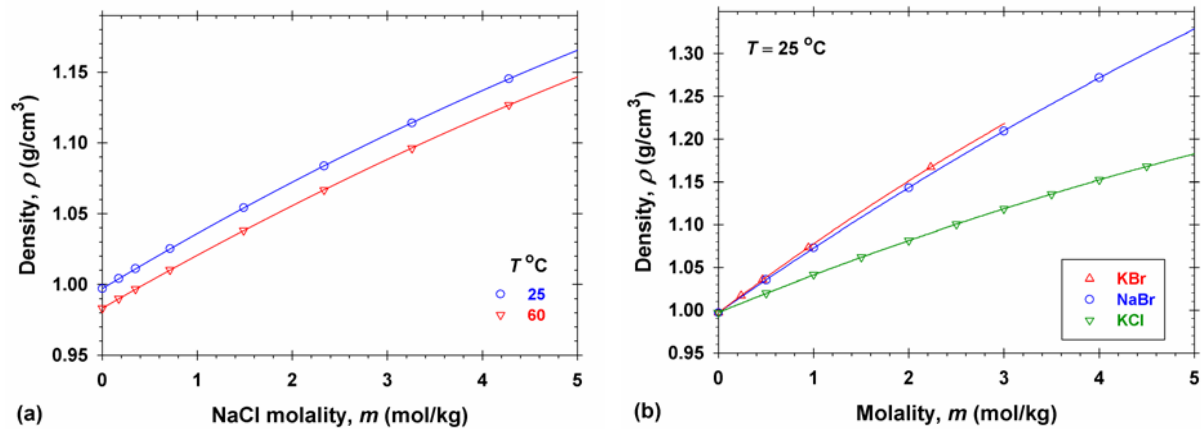

**Fig. G1**. Dependence of the solution's density $\rho$ on its molality $m$: (a) NaCl at 25 $^\circ$C and 60 $^\circ$C; (b) NaBr, KCl, and KBr at 25 $^\circ$C. The symbols are experimental data; the lines show the interpolation curves.